\documentclass{aa}  
\usepackage{subfigure}
\usepackage{threeparttable}
\usepackage{parskip}
\usepackage{float}
\usepackage{hyperref}
\usepackage{graphicx}
\usepackage{color}
\usepackage{url}
\usepackage[normalem]{ulem}
\usepackage{placeins} %
\usepackage{txfonts}

\newcommand{\nicer}{{\it NICER\/}} %

\newcommand{\xmm}{{\it XMM-Newton\/}}

\newcommand{\maxi}{{\it MAXI\/}} %

\newcommand{\rxte}{{\it RXTE\/}}

\def\be{\begin{equation}} 
\def\ee{\end{equation}}

\def\be{\begin{equation}} 
\def\ee{\end{equation}}

\usepackage{graphicx}
\usepackage{txfonts}
\begin{document} 
\title{NICER views moderate, strong, and extreme photospheric expansion bursts from the ultracompact X-ray binary 4U 1820--30}
   \subtitle{}

\author{Wenhui Yu\inst{1}
        \and
        Zhaosheng Li\inst{1}
        \thanks{Corresponding author}
        \and
        Yongqi Lu\inst{1}
        \and
        Yuanyue Pan\inst{1}
        \thanks{Corresponding author}
        \and
        Xuejuan Yang\inst{1}
        \and
         Yupeng Chen\inst{2}
        \and
         Shu Zhang\inst{2}
        \and
         Maurizio Falanga \inst{3,4}
         }
   \offprints{Z. Li}

   \institute{Key Laboratory of Stars and Interstellar Medium, Xiangtan University, Xiangtan 411105, Hunan, P.R. China\\ \email{lizhaosheng@xtu.edu.cn, panyy@xtu.edu.cn}
       \and
Key Laboratory of Particle Astrophysics, Institute of High Energy Physics, Chinese Academy of Sciences, 19B Yuquan Road, Beijing 100049, China
   \and
International Space Science Institute (ISSI), Hallerstrasse 6, 3012 Bern, Switzerland
           \and
Physikalisches Institut, University of Bern, Sidlerstrasse 5, 3012 Bern, Switzerland
}

   \date{Received XX; accepted XX}

  \abstract
  {Type I X-ray bursts in the ultracompact X-ray binary 4U 1820--30 are powered by the unstable thermonuclear burning of hydrogen-deficient material. We report the detection of 15 type I X-ray bursts from 4U 1820--30 observed by \nicer\ between 2017 and 2023. All these bursts occurred in the low state for the persistent flux in the range of $2.5-8\times10^{-9}~{\rm erg~s^{-1}~cm^{-2}}$ in 0.1--250 keV. The burst spectra during the tail can be nicely explained by blackbody model. However, for the first $\sim$5 s after the burst onset, the time-resolved spectra showed strong deviations from the blackbody model. The significant improvement of the fit can be obtained by taking into account of the enhanced persistent emission due to the Poynting-Robterson drag, the extra emission modeled by another blackbody component or by the reflection from the surrounding accretion disk. The reflection model provides a self-consistent and physically motivated explanation. We find that the accretion disk density changed with 0.5 s delay in response to the burst radiation, which indicates the distortion of the accretion disk during X-ray bursts. From the time-resolved spectroscopy, all bursts showed the characteristic of photospheric radius expansion (PRE). We find one superexpansion burst with the extreme photospheric radius $r_{\rm ph}>10^3$ km and blackbody temperature of $\sim 0.2$ keV, 13 strong PRE bursts for $r_{\rm ph}>10^2$ km, and one moderate PRE burst for $r_{\rm ph}\sim55$ km. } 

   \keywords{Stars: neutron - X-rays: bursts – accretion, accretion disks – X-rays: binaries--X-rays: individuals: 4U 1820--30
               }
\titlerunning{4U 1820--30}
\authorrunning{Yu et al.}
   \maketitle

\section{Introduction}
Ultracompact X-ray binaries (UCXBs) are a subtype of low-mass X-ray binaries (LMXBs), defined by their orbital periods of less than 80 minutes, in which a neutron star
(NS) or black hole  accretes matter from its low-mass companion \citep{Rappaport1982ApJ, 1995xrbi.nasa..457V}. During such short orbital periods, the donor star must be small enough to fit within the orbital size, such as white dwarfs or helium stars \citep{1981Paczynski, Nelson1986ApJ}. 
If the compact object is a NS, the accreted material is accumulated on the stellar surface which could occasionally trigger unstable thermonuclear bursts \citep[also referred to as type I X-ray burst, or simply X-ray burst, see e.g.,][for reviews]{Lewin93,Strohmayer06,Galloway21}.

The chemical composition of the accreted matter in LMXBs affects the properties of X-ray bursts \citep{Nathalie2018Accretion}. The burst duration depends on the cooling of the burning layer, and therefore scales with the ignition depth \citep{Cumming2004ApJ,int2014A&A}. Normal type I  X-ray bursts are fueled by a mixture of hydrogen and helium or pure helium, lasting $\sim$10--100 seconds with a recurrence time of hours \citep{Lewin93}. Intermediate-duration bursts are due to unstable burning of pure helium in deeper layers \citep{Cumming2006ApJ}, with a duration of tens of minutes and a recurrence time of several weeks \citep{Falanga2008A&A,int2011A&A,int2019A&A,bult2021}. Superbursts are powered by carbon at an ignition depth  of $\sim10^{11-12} \mathrm{~g~ cm^{-2}}$ \citep{Strohmayer_2002,Cumming2003,keek2012ApJ,int2017}, and last from hours to days and recur from days to years \citep{keek2015,int2019A&A,li2021ApJ}. The X-ray burst spectra can be described by a blackbody model with temperatures in the range of $kT_{\rm bb}\approx0.5-3$ keV.    In UCXBs, resulting from the unstable burning of helium-rich gas accreted from the companion, the burst peak fluxes usually reach or slightly exceed the Eddington limit, showing the photospheric radius expansion (PRE) from time-resolved spectroscopy  \citep[][]{Lewin93}. Moreover, some bursts exhibit strong expansion of the radius excess $\sim$100 km \citep{Galloway08}. A small fraction bursts are so extreme that the radii reach $\sim 10^3 \rm~km$, implying the photosphere expands a factor of > 100, which is referred to as a superexpansion burst \citep{int2010,int2012A&A}.

The energetic X-ray photons released from the bursts can interact with the matter surrounding the NS in various ways, and can significantly impact both the disk and corona \citep{int2011A&A,Degenaar18,int2019A&A,Fragile2020}.  Numerical simulation reveals that the outpouring of radiation during an X-ray burst affects the disk structure and  generates radiatively driven warps in a large fraction of the disk \citep{Ballantyne2023}. Observationally, the evidence is usually found in the emergent spectra of X-ray bursts that deviate from the Planck function. The enhanced persistent emission during the X-ray bursts caused by the Poynting-Robertson drag has been proposed to produce the soft and hard excess in the burst spectra \citep{Zand13,Worpel13,Guver22}. Alternatively, the excess can also be explained by the burst emission that was reflected by the accretion disk \citep{Zhao22,Lu2023A&A,Speicher2022}.

4U 1820--30 is a persistent and atoll X-ray source, that was first discovered by the Uhuru satellite \citep{uhuru_1974ApJS}. It is also the first X-ray burster located in the globular cluster NGC 6642 at a distance of $8.4\pm{0.6}$ kpc \citep{Valenti2004}. The binary orbital period of 685 s suggests it is an UCXB \citep{Stella1987}.  The donor of 4U 1820--30 is probably a helium white dwarf with a mass of $\sim 0.06 - 0.07 ~M_{\odot}$ \citep{Rappaport1987}. The binary inclination angle to the observer is constrained in the range of $35^{\circ}-50^{\circ}$ \citep{Anderson_1997}. The source only  shows X-ray bursts during the low state with the recurrence time of 2--4 hr \citep{Clark76,int2012A&A,garc2013,Suleimanov2017}. Two superbursts and two long X-ray bursts  have been observed in 4U 1820--30 \citep{Strohmayer_2002,int2011ATel,Serino2016,a2021ATel,b2021ATel}.    Comparing the thermonuclear flash models with the observed recurrence time, burst fluences and persistent fluxes, \citet{Cumming2003} proposed that a small fraction of hydrogen and heavy elements is required  in accreted matter. We note that \citet{Keek2018} reported a strong PRE burst in 4U 1820--30 from \nicer\ observations. In addition to a blackbody component, an extra optically thick Comptonization model has to be added to fit the time-resolved burst spectra appropriately. \citet{Strohmayer2019} found the emission line at 1 keV, and the absorption lines at 1.7 and 3.0 keV, during the PRE phases during X-ray bursts in 4U 1820--30.

Beyond the analysis of \citet{Keek2018},  we study all X-ray bursts from 4U 1820--30 detected by \nicer between 2017 and 2023 in this work. In Section \ref{Sec:observe}, we describe the properties of the persistent emissions and X-ray bursts in between 2017-2023. In Section \ref{sec:spec_analysis}, we perform the spectral fitting of the persistent emissions and the time-resolved spectra of the X-ray bursts. We discuss and summarize the results in Section \ref{Sec:discussion} and Section~\ref{Sec:conclusion}, respectively.

\section{Observations and data reduction}
\label{Sec:observe} 
\nicer\ has observed 4U 1820--30  with the net unfiltered exposure time of 244, 8, 149, 17, 141, 166, and 29 ks in 2017--2023, for a total exposure time of 754 ks.  We processed all the archived \nicer\ data by applying the standard filtering criteria using HEASOFT V6.30.1 and the \nicer\ Data Analysis Software (NICERDAS) V1.0.2.  The 1 s binned light curves were extracted  in the energy ranges between 0.5--10 keV from the calibrated unfiltered (UFA) files and cleaned events files to search for type I X-ray bursts.  We found 15 type I X-ray bursts, 5 in 2017, 9 in 2019, and 1 in 2021. Three of them are only shown in the UFA files (see Table~\ref{table:burst_obs}). The spectral results of  bursts \#1 and \#1--5 are reported in \citet{Keek2018} and \citet{Strohmayer2019}, respectively.  It is worth noting that the \nicer\ data did not cover two long X-ray bursts observed by \maxi. We also extracted 64 s binned light curves in the energy ranges of 0.5--1.1, 1.1--2.0, 2.0--3.8, and 3.8--6.8 keV, using the command ${\tt xselect}$. We calculated the soft and hard colors as the count rate ratios of  1.1--2.0 keV/0.5--1.1 keV and 3.8--6.8 keV/2.0--3.8 keV, respectively. The intensity is defined as the count rate covering the energy range of 0.5--10 keV. 

\begin{table*}
\begin{center} 
\caption{Observation IDs and light curve properties of all 15 type I X-ray bursts.  \label{table:burst_obs}}
\begin{tabular}{ccccccc}
\hline\\ %
{\centering  Burst No. } &
{\centering \nicer} &
{\centering  Burst Onset  } &
{\centering  Peak Rate\tablefootmark{c}} &
{\centering  dip duration\tablefootmark{d}} &
{\centering  dip count\tablefootmark{e}} &\\
$\#$ &(Obs Id)& (MJD) &($\mathrm{10^{4} ~ c ~ s^{-1}}$)&(s)& \% &\\ [0.01cm] \hline
1  & 1050300108    &57994.37393&2.15&$0.6\pm{0.1}$&$47.5\pm{0.1}$  &\\
2  & 1050300108    &57994.46448&1.56&-            &-              &\\
3  & 1050300109    &57995.22557&2.28&$0.7\pm{0.1}$&$31.3\pm{0.1}$  &\\
4  & 1050300109  &57995.34162&2.08&$0.7\pm{0.1}$&$41.4\pm{0.1}$  &\\
5  & 1050300109    &57995.60600&1.89&$1.0\pm{0.1}$&$61.1\pm{0.1}$  &\\
6  & 2050300104               &58641.29573&2.52&$1.9\pm{0.1}$&$8.4 \pm{0.1}$  &\\
7  & 2050300108               &58646.84015&1.96&$2.0\pm{0.1}$&$-0.9 \pm{0.1}$ &\\
8  & 2050300110               &58648.32669&2.25&$2.7\pm{0.1}$&$-1.0 \pm{0.1}$ &\\
9  & 2050300115\tablefootmark{a}&58655.09203&2.59&$1.7\pm{0.1}$&$15.5\pm{0.1}$  &\\
10 & 2050300119\tablefootmark{b}    &58660.31327&2.60&$1.7\pm{0.1}$&$8.3 \pm{0.1}$  &\\
11 & 2050300119               &58660.77147&2.58&$1.5\pm{0.1}$&$17.7\pm{0.1}$  &\\
12 & 2050300120\tablefootmark{a}&58661.78614&2.22&$1.0\pm{0.1}$&$27.5\pm{0.1}$  &\\
13 & 2050300122\tablefootmark{a}&58663.98438&2.57&$1.1\pm{0.1}$&$20.4\pm{0.1}$  &\\
14 & 2050300124               &58665.59405&2.54&$1.6\pm{0.1}$&$13.4\pm{0.1}$  &\\
15 & 4680010101               &59336.61230&2.57&$1.7\pm{0.1}$&$13.6\pm{0.1}$  &\\

\hline 
\end{tabular} 
\end{center}

\tablefoot{
Bursts \#1 and \#1-5 have been studied in \citet{Keek2018} and \citet{Strohmayer2019}, respectively.\\
\tablefoottext{a} {The X-ray bursts in the ufa event file.}\\
\tablefoottext{b}{ The tail of burst \#10 was truncated due to data gap.}\\
\tablefoottext{c}{ The peak rates with persistent emission subtracted are measured from the 0.1 s light curves in the energy range of 0.5--10 keV.}\\
\tablefoottext{d} {The dip duration of the 0.1 s light curves in the energy range of 3--10 keV.}\\
\tablefoottext{e}{ The ratio of minimum count rate during the dip to peak count rate in 3--10 keV.}
}
\end{table*}

\begin{figure*}
    \sidecaption
        \includegraphics[width=\hsize]{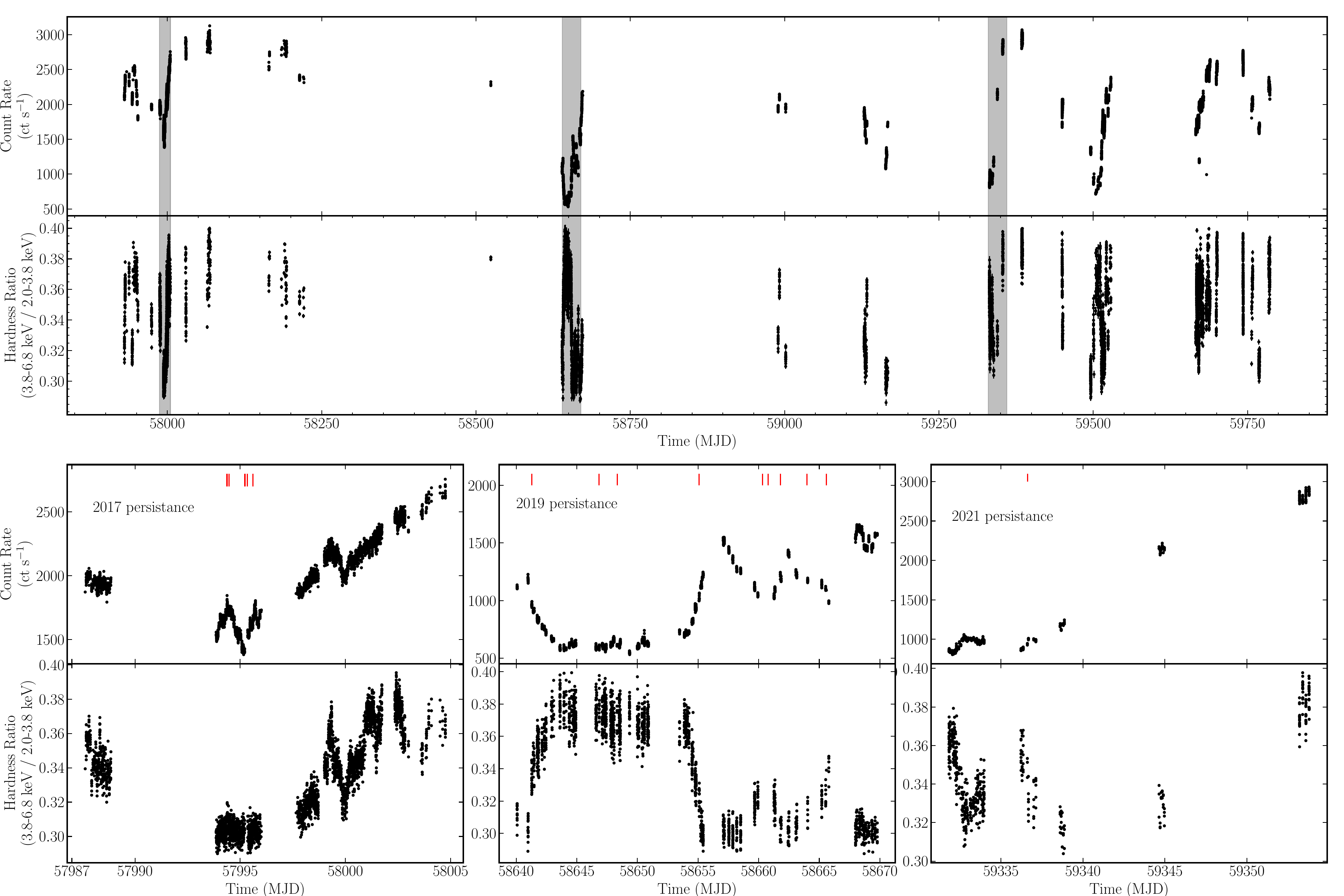}
\caption{ Light curve and hardness of 4U 1820--30 observed by \nicer\ from 2017 to 2022. The top panel shows the light curve in the 0.5--10 keV and the hardness ratio between 3.8--6.8 keV and the 2.0--3.8 keV. Each black point means 64 s of data from \nicer\, and all bursts are removed. The gray intervals cover the X-ray bursts, which correspond to three panels at the bottom. In the bottom panel, from left to right, are plotted the persistent emissions and the hardness ratios  between MJD $57987-58005$, $58640-58670$, and $59330-59360$. We indicate the onset of all bursts with red bars. 
}
\label{Fig:pre_burst}
\end{figure*}

\subsection{The persistent light curves}
\label{Sec:spec_per}
The light curves and hardness ratio of 4U 1820--30 observed by \nicer\ from June 2017 to  August 2023 are shown in the top panel of Fig.~\ref{Fig:pre_burst}. The gray regions in the top panels covere the epoch of the appearance of X-ray bursts; zoomed in images are shown in the bottom panels. The bottom left panel of Fig.~\ref{Fig:pre_burst} shows the 2017 pre-burst persistent flux observed by \nicer\ between MJD $57987-58005$. The pre-burst persistent light curve slowly decreased  from $\sim2000$ to  $\sim1400$ ${\rm c~s^{-1}}$ in  five days before the onset of the first X-ray burst in our sample. It fluctuated between $\sim1400$ and $\sim1800$ ${\rm c~s^{-1}}$ within two days when the burst was observed, and after that it rose to $\sim3000$ ${\rm c~s^{-1}}$ within ten days. The trends of the hardness ratio  during this period were similar to the light curve, decreasing from 0.38 to 0.3 in the first five days, and then rising to 0.4 in a relatively short time. The bottom conter panel of Fig.~\ref{Fig:pre_burst} presents the 2019 observations between MJD $58640-58670$, and the light curve decreased from $\sim1200$ to $\sim600$ ${\rm c~s^{-1}}$ in the first five days. After
maintaining a low count rate for the next ten days, it rapidly increased to $\sim1600$ ${\rm c~s^{-1}}$ in two days, and varied in the subsequent 15 days. However, the trends of hardness ratio during this period were different from that of 2017, which showed the opposite trend compared with the light curve. The bottom right panel of Fig.~\ref{Fig:pre_burst} displays the 2021 observations between MJD $59330-59360$, the light curve was slowly rising when the burst was observed, and the hardness ratio  decreased first and then increased to around 0.4. 

\begin{figure}
   \sidecaption
   \includegraphics[width=9cm]{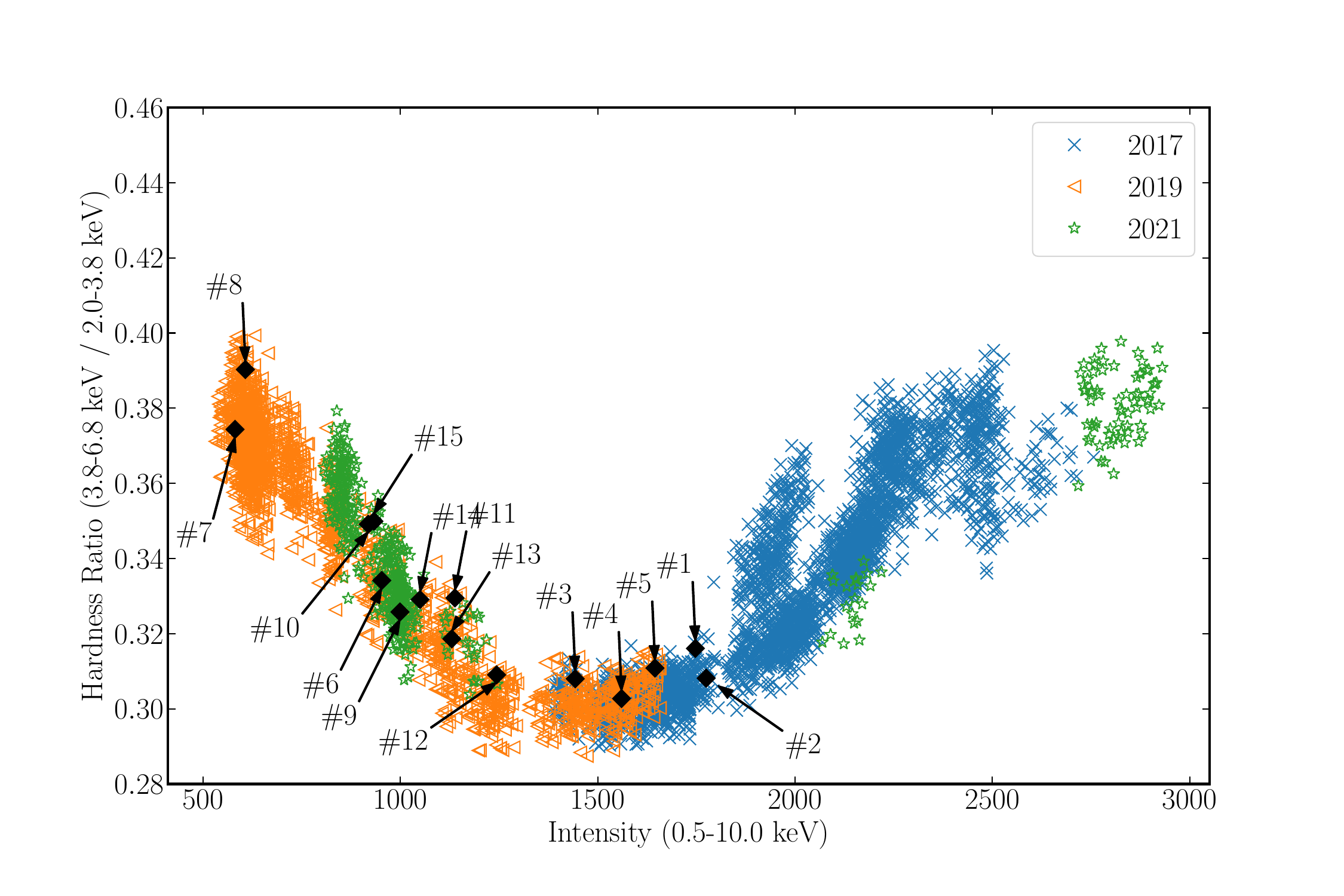}
   \includegraphics[width=9cm]{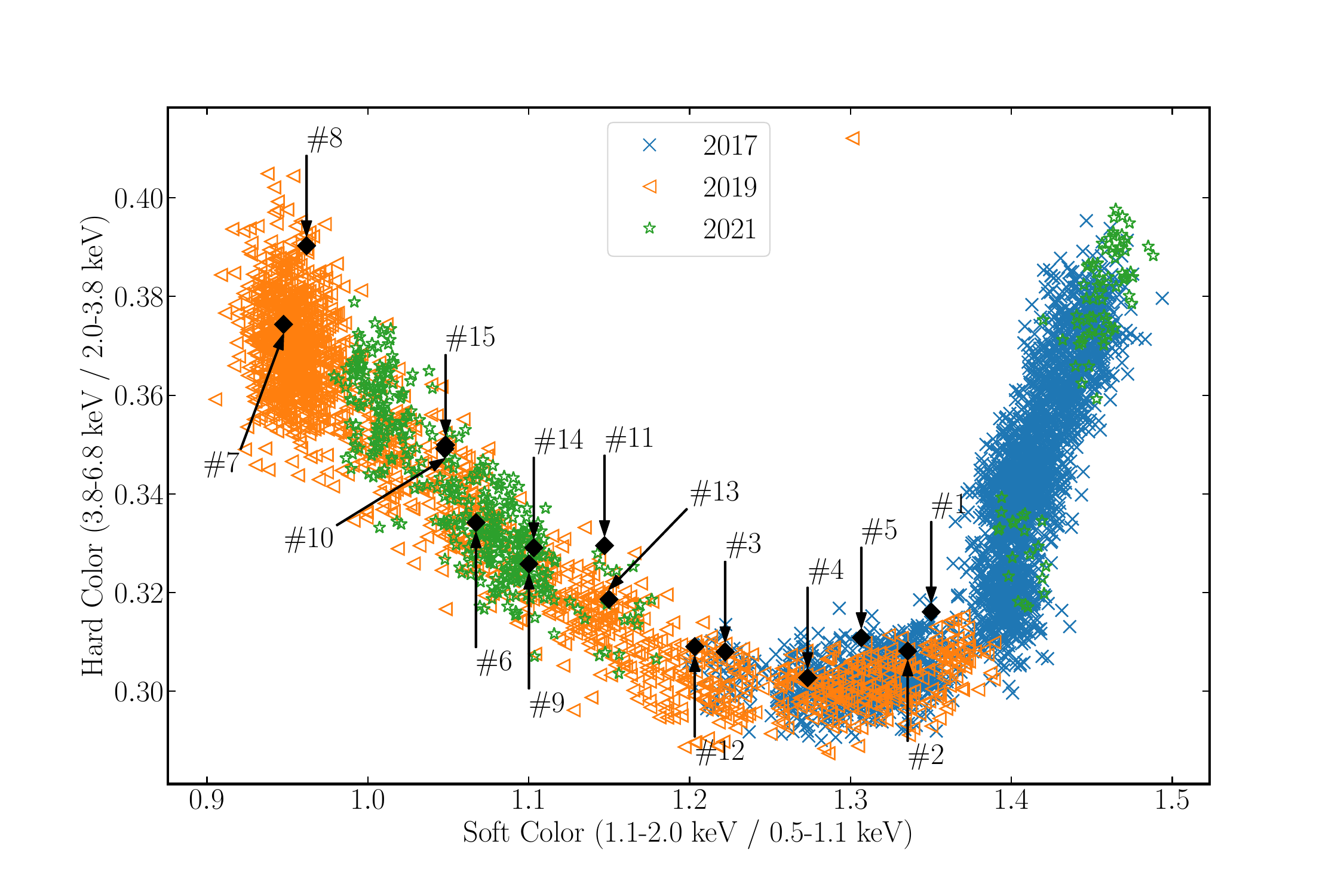}
      \caption{Hardness-intensity diagram (HID, {\it Top panel}) and the color–color diagram (CCD, {\it  Bottom panel}) of 4U 1820--30 from the \nicer\ observations. The blue crosses, orange triangles and green stars represent the data observed in 2017, 2019, and 2021, respectively. All bursts are removed, and each point represents a segment of 64 s. The HID and CCD of the persistent emission before each X-ray burst are shown as filled black diamonds. For each panel, from left to right, the spectra evolved from the hard island state to the soft banana state.
              }
\label{fig:HID}
\end{figure}

The hardness intensity diagram (HID) and the color–color diagram (CCD) of the 2017, 2019 and 2021 observations from 4U 1820--30 are displayed in Fig.~\ref{fig:HID}. Compared with the results from \rxte\ \citep[see Fig. 4 in][]{int2012A&A}, although the CCD produced in different energy bands, its shape from the \nicer\ observations is quite similar, but rotates around 90 degrees counterclockwise. The data have two branches, the high (soft banana) state  and low (hard island) state. The HID and CCD prior to each X-ray burst are also marked in Fig.~\ref{fig:HID}. There no bursts are detected for the soft color > 1.38 or the intensity > 1800 ${\rm c~s^{-1}}$. 
It is consistent with a well-known fact that the bursts in 4U 1820--30 took place predominantly at low luminosity \citep{int2012A&A}.

\subsection{The X-ray bursts light curves}
\label{Sec:burst_lc}

\begin{figure*}
\includegraphics[width=\hsize]{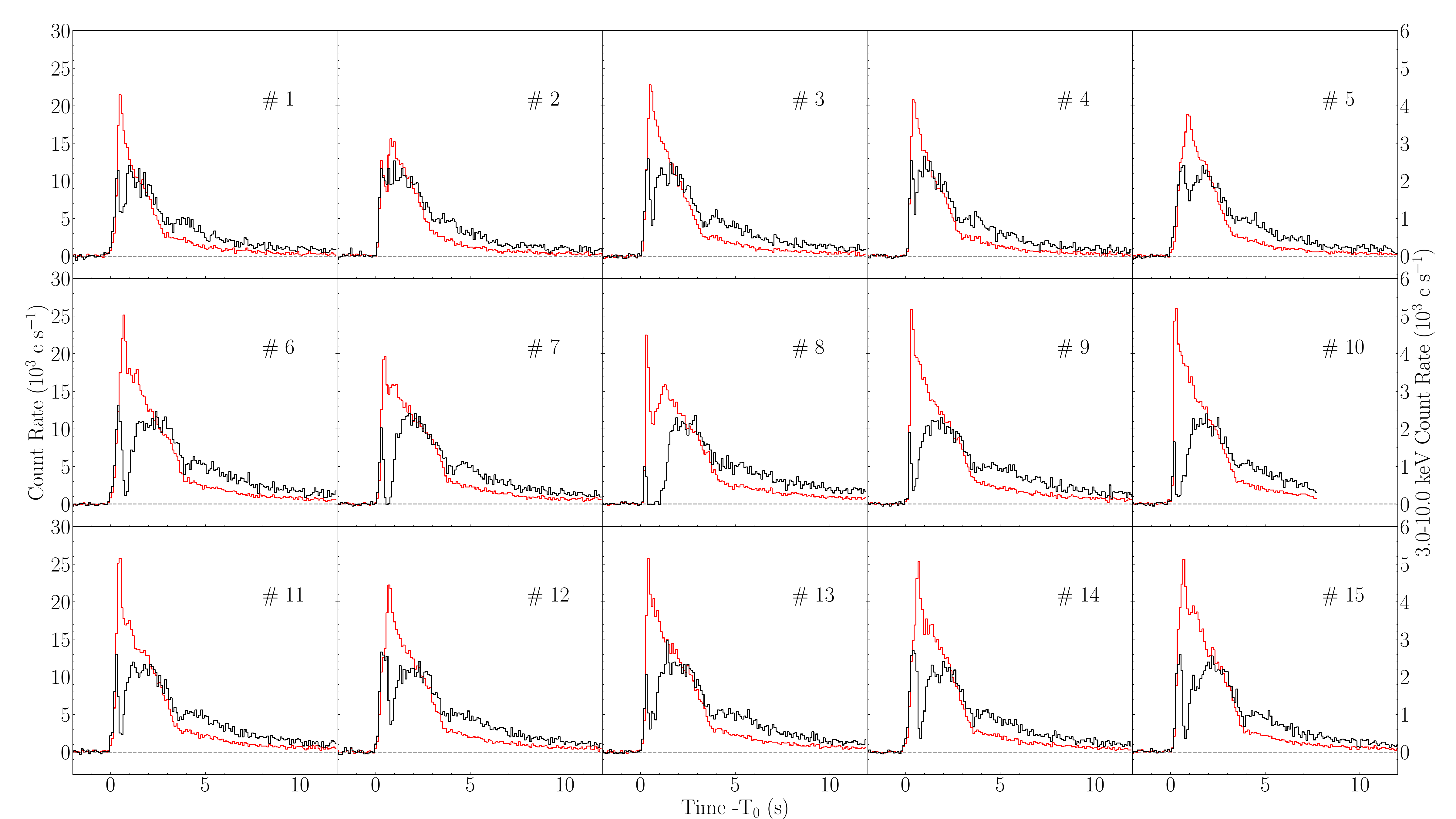}
   \caption{0.1 s binned light curves (red lines) of 15 X-ray bursts from 4U 1820--30 observed by \nicer\ in 0.5--10 keV. The right axis shows the 3--10 keV light curves as black lines. The persistent emissions are regarded as background and were subtracted. }
      \label{fig:lc}
\end{figure*}

We extracted the 0.1 s binned light curves for all 15 bursts in two energy bands, 3--10 keV and  0.5--10 keV. The burst onset time is set as the count rate that exceeds the pre-burst rate by a factor of 1.5 from the 0.5--10 keV band. The pre-burst rate  is defined as an average count rate for each band in a 64 s window ending 10 s prior to the burst onset time. The burst end time is determined as the point where the burst count rate has decayed to the pre-burst level. The burst epoch MJD time and the peak rate in 0.5--10 keV of all 15 bursts are listed in Table~\ref{table:burst_obs}.  The corresponding light curves in two energy bands are displayed in Fig. \ref{fig:lc}. A part of the tail in burst \#10 was truncated due to the data gap. The source persistent emission is considered to be background and was subtracted from the burst count rate for each band.  Most of the bursts have the peak count rates in the range of $1.9-2.5\times10^4~{\rm c~s^{-1}}$, while burst \#2 has a slightly lower count rate with $\sim1.5\times10^4~{\rm c~s^{-1}}$. 
The X-ray burst light curves in 0.5--10 keV rise rapidly in 0.2--1 s and decrease slowly in $\sim10$ s, which is consistent with a hydrogen-poor burst.

In Fig.~\ref{fig:lc},  dips obviously appear in bursts \#2 and \#8 around their peaks in 0.5--10 keV. The light curve shapes in 3--10 keV are quite different with the 0.5--10 keV. The tails in 3--10 keV light curves show a second peak rather than an exponential decay. Moreover, dips in 3--10 keV are visible in all bursts but \# 2.  The duration of the dip is defined as the time between the first peak and the following peak of hard light. The dip count is the ratio of minimum photon rate during the dip to peak photon rate in 3--10 keV. Except for burst \#2, the bursts exhibit a dip around the burst peak which lasted 0.6--2.7 s, similar to the bursts observed by \rxte\ for the comparable energy range \citep{int2012A&A}. It is also found that the dip count rate in 3--10 keV in bursts \#7 and \#8 dropped to below their pre-burst levels.

Since the spin frequency of 4U 1820--30 is unknown, we searched for  the burst oscillation from the event files.  The power spectra is calculated in 0.3--12, 0.3--3, 3--6, and 6--12 keV, by applying the Fast Fourier transform (FFT) with Leahy normalization \citep{Leahy83}. We adopt the moving window method with the width of $\Delta T=4$ s, and steps of 0.5 s. For each window, the independent Fourier frequencies between 50 and 2000 Hz and the corresponding power are computed with steps of 0.25 Hz \citep[see, e.g.,][]{Bilous19, li2022ApJ}. No significant burst oscillations were detected in these bursts.   The fractional rms amplitude is estimated from the relation, $A_{\rm {rms}} \approx \sqrt{P_{\rm s}/N_{\rm m}}$, where $N_{\rm m}$ is the total X-ray photons of the burst interval and $P_{\rm s}$ is the maximum power of the signal. We obtained the upper limit of the fractional rms amplitude to be $5.7\%$.

\section{Spectral analysis}   
\label{sec:spec_analysis}
We performed the spectra analysis using Xspec v12.12.0 \citep{Arnaud96}. All spectral models mentioned in this work can be found in the Xspec manual webpage.\footnote{\url{https://heasarc.gsfc.nasa.gov/xanadu/xspec/manual/manual.html}}  We extracted the spectra and the 3c50 background spectra  \citep[][]{Remillard21}. The associated ancillary response files ($\texttt{ARFs}$) and response matrix files ($\texttt{RMFs}$) were produced. The uncertainties of all parameters are quoted at the $1\sigma$ confidence level.

\subsection{Persistent spectra}
\label{Sec:spec_per}
We extracted the pre-burst persistent spectra in a 64 s window ending 10 s prior to the burst onset time for all 15 X-ray bursts in 4U 1820--30. We performed the optimal binning for each persistent spectrum by using \texttt{ftgrouppha} as recommended by the \nicer\ team. Following the broadband spectral analysis by \citet{Costantini2012}, we fitted the pre-burst persistent spectra with a combination of a blackbody (\texttt{bbodyrad} in Xspec) and a Comptonization component \citep[\texttt{compTT};][]{Lev}, modified by the Tübingen–Boulder model, ${\tt TBabs}$, with abundances from \citet{Wilms_2000}. These two components represent the thermal radiation from the accretion disk (NS and/or boundary layer) and the Comptonization from the corona around the disk, respectively. The free parameters are, the equivalent hydrogen column, $N_{\rm H}$, for ${\tt TBabs}$; the blackbody temperature, $kT_{\rm bb}$, and the normalization, $K_{\rm bb}$, for \texttt{bbodyrad}; the temperatures of the seed photons and hot electrons, $T_{\rm 0}$ and $T_{\rm e}$, respectively; and the optical thickness of the electron slab, $\tau_{\rm e}$, and the normalization, $K_{\rm comp}$, for \texttt{compTT}. The  \texttt{compTT} geometry was set as disk. We performed the joint spectral fitting in three groups; five pre-burst spectra in 2017, nine in 2019, and one in 2021. For each group, we tied the absorption column density across all spectra, and set $N_{\rm H}$, the parameters of the \texttt{bbodyrad} and \texttt{compTT} free to change.

This model can fit all persistent spectra well for the $\chi^2$ per degree of freedom (dof), $\chi^2_\nu$, close to unity. The best-fit parameters are listed in Table~\ref{table:preburst}. We obtained the absorption column density  $(2.33 \pm{0.04})\times 10^{21}$, $(1.93 \pm{0.07})\times 10^{21}$, and $(1.93 \pm{0.13}) \times 10^{21}~ {\rm cm}^{-2}$  for the pre-burst spectra in 2017, 2019 and  2021, respectively.  They are consistent with $N_{\rm H}=(2.5 \pm{0.3})\times 10^{21}$ and $(1.63 \pm{0.02}) \times 10^{21}~ {\rm cm}^{-2}$ obtained from the {\it Chandra} spectra \citep{Guver2010} and the {\xmm} spectra  \citep{Costantini2012}, respectively. No distinct features are appearent in the residual.
 We also calculated the unabsorbed bolometric flux in energy range of 0.1--250 keV by using the tool \texttt{cflux}.

\begin{figure}
\centering
\includegraphics[width=9cm,]{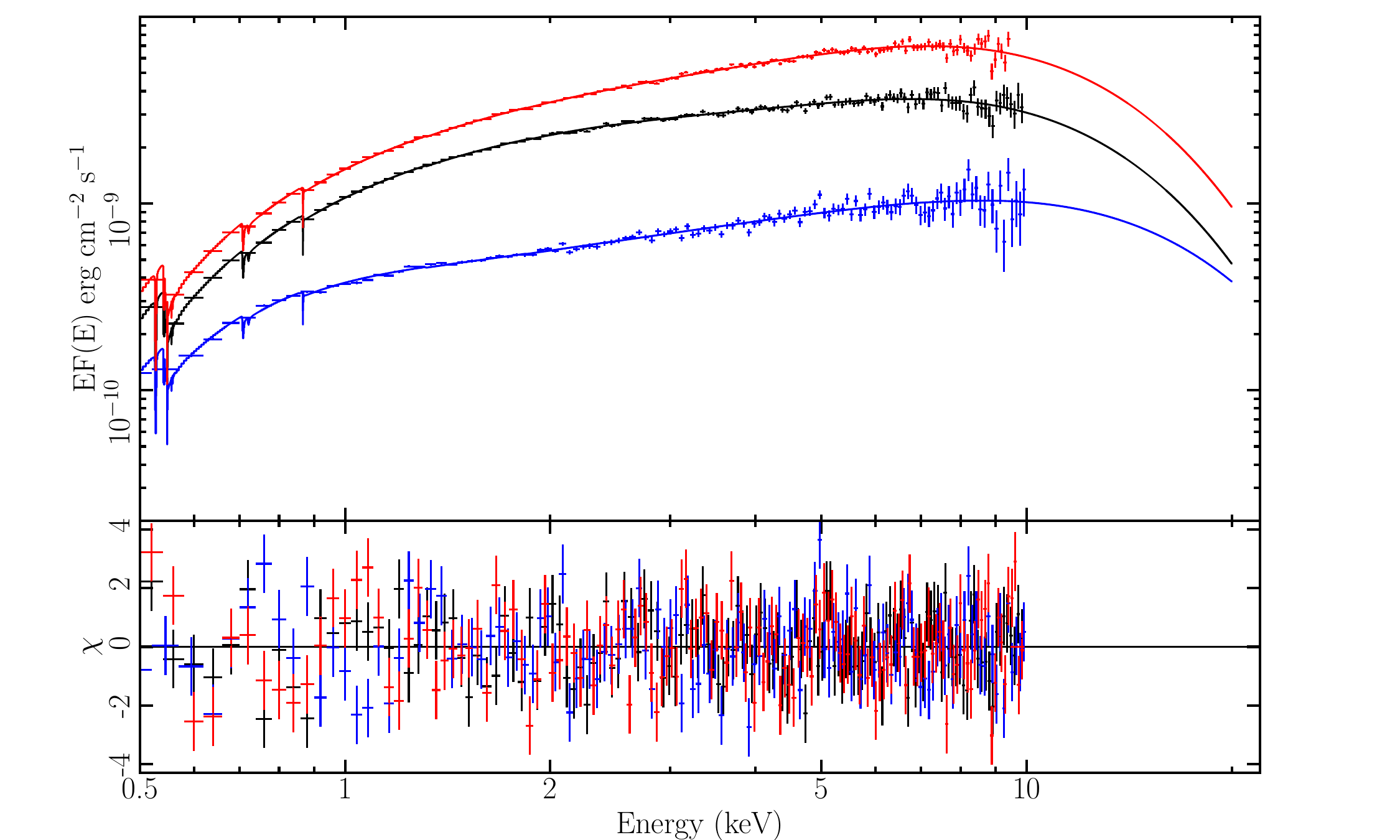}
\caption{Persistent spectra and their best-fit models of  4U 1820--30. {\it Top panel} (from top to bottom): Persistent spectra  from the right and middle of HID, and prior to burst \#8. The best-fit models, ${\tt Tbabs} \times ({\tt bbodyrad + compTT })$, are represented as solid lines for each spectrum.  {\it Bottom panel}: Residuals of the best-fit models to the spectra. }

\label{fig:eeuf}
\end{figure}

Moreover, we also generated tow other persistent spectra for the total exposure of 256 s, from the right of HID with the count rate of $\sim3000~{\rm c~s^{-1}}$ and hardness ratio of $\sim0.38$, and from the middle of HID with the count rate of $\sim2000~{\rm c~s^{-1}}$ and hardness ratio of $\sim0.32$. The above-mentioned model can also fit the spectrum from the middle of HID very well. The best-fitparameters are $N_{\rm H} = (2.40 \pm{0.11}) \times 10^{21}~ {\rm cm}^{-2}$, $kT_{\rm bb} = 0.55 \pm{0.03} $ keV, $R_{\rm bb} = 18 \pm{9}$ km, $T_{\rm 0} = 0.02 \pm{0.01}$ keV, $T_{\rm e} = 2.31 \pm{0.12}$ keV, $\tau_{\rm e} = 8.73 \pm{0.43}$, the unabsorbed bolometric flux in 0.1--250 keV, $F = (9.90 \pm{0.02}) \times 10^{-9}\enspace \mathrm{erg\enspace s^{-1}\enspace cm^{-2}}$, and $\chi^{2} = 147.6$ for 132 dof.
For the spectrum from the right of HID, the best-fit parameters of the continuum model are, $N_{\rm H} = (2.44 \pm{0.09}) \times 10^{21}~ {\rm cm}^{-2}$, $kT_{\rm bb} = 0.59 \pm{0.09} $ keV, $R_{\rm bb} = 15 \pm{9}$ km, $T_{\rm 0} = 0.06 \pm{0.02}$ keV, $T_{\rm e} = 2.3 \pm{0.1}$ keV, $\tau_{\rm e} = 9.5 \pm{0.4}$, the unabsorbed bolometric flux $F = (14.99 \pm{0.06}) \times 10^{-9}\enspace \mathrm{erg\enspace s^{-1}\enspace cm^{-2}}$, and $\chi^{2} = 229.7$ for 136 dof. In Fig. \ref{fig:eeuf} we show these two spectra and the pre-burst persistent spectrum from burst \#8. The best-fit models are also shown.

From high to low state, the unabsorbed bolometric flux decreased from $15 \times 10^{-9}$ to $2.5\times 10^{-9}\enspace \mathrm{erg\enspace s^{-1}\enspace cm^{-2}}$, and  the absorption column density also dropped from $2.5\times 10^{21}$ to $1.9\times 10^{21}~ {\rm cm}^{-2}$.  In the low state, the blackbody component covered the a lowest temperature, $\sim$ 0.1 keV, with a radius around 100 km. As the spectra move to the high state, the blackbody temperature  increased from 0.2 to 0.6 keV and its radius shrink from $\sim$100 to $\sim$14 km. The size of the blackbody component of all the spectra are larger than the typical radius of NS, which implies its emission from the accretion disk. Two spectral lines were only shown in the extreme high state. 
The plasma temperature is very close across these spectra in different states. Compared with the low state, the optical depth is higher in the high sate. All bursts occurred at the persistent flux $\approx2.5-8\times10^{-9}~\rm{erg~s^{-1}~cm^{-2}}$ in 0.1-250 keV, that is $\sim5-15\%~L_{\rm Edd}$. At higher flux, the activity of X-ray bursts was quenched.

\begin{table*}[h!]
\begin{center} 
\caption{Joint-fit pre-burst spectral parameters.  \label{table:preburst}}
\resizebox{\linewidth}{!}{\begin{tabular}{ccccccccccccc} 
\hline\\ %
{\centering  Burst  } &
{\centering  $N_{\rm H}$  } &
{\centering  $kT_{\rm BB}$} &
{\centering  $R_{\rm BB}$} &
{\centering  ${kT_{0}}$} &
{\centering  $kT_{\rm e}$\tablefootmark{a}} &
{\centering  $\tau$} &
{\centering  $\chi^{2}_{\rm red}$}&
{\centering  ${F_{\rm per}}$\tablefootmark{b}} &\\
$\#$ & $(10^{21}\enspace  \mathrm{cm^{-2}})$ &$\mathrm{(keV)}$ & $(\mathrm{km})$ &$\mathrm{(keV)}$& $\mathrm{(keV)}$& && ($10^{-9}~\mathrm{erg~ s^{-1}~cm^{-2}}$)&\\ [0.01cm] \hline
1  &                &$0.46\pm{0.05}$&$30\pm{18}$ &$0.043\pm{0.006}$& $2.16\pm{0.02}$    &$8.65\pm{0.05}$       & 1.09(120)& $7.84\pm{0.09}$& \\
2  &                &$0.49\pm{0.05}$&$28\pm{17}$ &$0.010\pm{0.009}$& $2.49\pm{0.01}$    &$7.33\pm{0.03}$       & 1.16(121)& $8.02\pm{0.09}$& \\
3  & 2.33$\pm{0.03}$&$0.41\pm{0.03}$&$34\pm{19}$ &$0.078\pm{0.001}$& $3.15\pm{0.02}$    &$6.45\pm{0.05}$       & 1.11(120)& $6.48\pm{0.09}$& \\
4  &                &$0.42\pm{0.03}$&$35\pm{20}$ &$0.043\pm{0.006}$& $2.86\pm{0.02}$    &$7.03\pm{0.04}$       & 1.10(120)& $7.23\pm{0.09}$& \\
5  &                &$0.43\pm{0.03}$&$35\pm{19}$ &$0.073\pm{0.001}$& $2.61\pm{0.02}$    &$7.58\pm{0.06}$       & 1.03(121)& $7.29\pm{0.09}$& \\\hline
6  &                  &$0.32\pm{0.03}$&$47\pm{23}$&$0.097\pm{0.016}$&$5.01_{-1.76}^{+p}$ &$5.21\pm{1.43}$       & 1.09(112)& $4.44\pm{0.08}$& \\
7  &                  &$0.18\pm{0.03}$&$95\pm{59}$&$0.023\pm{0.019}$&$2.32\pm{0.46}$     &$8.23\pm{0.75}$       & 1.12(109)& $2.52\pm{0.09}$& \\
8  &                  &$0.20\pm{0.06}$&$81\pm{41}$&$0.071\pm{0.020}$&$2.84\pm{0.51}$     &$7.31\pm{1.13}$       & 1.03(109)& $2.56\pm{0.08}$& \\
9  &                  &$0.33\pm{0.03}$&$44\pm{18}$&$0.101\pm{0.005}$&$3.23\pm{0.49}$     &$6.50\pm{1.41}$       & 1.14(113)& $4.09\pm{0.09}$& \\
10 & $1.93\pm{0.07}$  &$0.29\pm{0.02}$&$55\pm{10}$&$0.071\pm{0.001}$&$3.17\pm{0.45}$     &$6.62\pm{1.18}$       & 1.15(113)& $3.91\pm{0.09}$& \\
11 &                  &$0.35\pm{0.02}$&$49\pm{8}$ &$0.092\pm{0.001}$&$2.91\pm{0.51}$     &$7.20\pm{1.91}$       & 1.06(115)& $4.66\pm{0.07}$& \\ 
12 &                  &$0.37\pm{0.05}$&$47\pm{9}$ &$0.082\pm{0.002}$&$2.08\pm{0.33}$     &$8.86\pm{2.81}$       & 1.04(103)& $4.91\pm{0.06}$& \\
13 &                  &$0.35\pm{0.02}$&$45\pm{22}$&$0.099\pm{0.013}$&$2.87\pm{0.58}$     &$7.18\pm{1.22}$       & 1.04(115)& $4.57\pm{0.09}$& \\
14 &                  &$0.32\pm{0.01}$&$54\pm{21}$&$0.002\pm{0.001}$&$10.35_{-1.22}^{+p}$&$1.15^{+4.25}_{-0.01}$& 0.95(114)& $5.39\pm{0.09}$& \\\hline
15 & 1.93$\pm{0.13}$&$0.28\pm{0.02}$&$46\pm{+21}$ &$0.037\pm{0.036}$&$7.13_{-3.93}^{+p}$ &$4.34\pm{2.32}$       & 1.08(112)& $4.90\pm{0.07}$& \\ 
\hline 
\end{tabular} }
\end{center}

\tablefoot{
\tablefoottext{a}{ The symbol $p$ means that the poorly constrained electron temperature is pegged at the hard limit.}\\
\tablefoottext{a}{ The unabsorbed bolometric persistent flux in the energy range of $0.1-250$ keV.}
}

\end{table*}   

\subsection{X-ray burst time-resolved spectroscopy}
\label{Sec:spec_burst}
To investigate the time-resolved spectroscopy during X-ray bursts, we extracted the burst spectra with the exposure time varying between 0.1 and 4 s to guarantee that each spectrum had at least 1500 counts in the energy range 0.5--10 keV. All burst spectra were grouped using ${\tt grappha}$ with a minimum count of 20 in each bin. We used an absorbed blackbody model, \texttt{TBabs} $\times$ \texttt{bbodyrad}, to fit the burst spectra, where the pre-burst spectrum was regarded as background and unchanged during bursts. We fixed the hydrogen column density at $2.33 \times 10^{21}$,  $1.93 \times 10^{21}$, and $1.93 \times 10^{21}~{\rm cm} ^{-2}$, for the X-ray bursts observed in 2017, 2019, and 2021, respectively.  We found that the blackbody model can fit  the spectra  well during the cooling tail. However, for the first $\sim$ 5 s of all 15 X-ray bursts, the model fitted the spectra  poorly (i.e., the maximum $\chi^2_\nu$ $\sim3-6$).  The fit can be significantly improved by using the enhanced persistent emission model (Sec.~\ref{Sec:spec_fa}), adding a reflection component from the surrounding accretion disk (Sect.~\ref{Sec:spec_diskref}), or the double-blackbody model (Sect.~\ref{Sec:spec_2bb}). In Sects.~\ref{Sec:spec_fa}, \ref{Sec:spec_diskref}, and \ref{Sec:spec_2bb}, the absorption and/or emission lines were also involved if necessary, but  the  details will be provided in as separate paper.  

\subsubsection{Enhanced persistent emission}
\label{Sec:spec_fa}
We used the $f_{\rm a}$ model, \texttt{TBabs} $\times$ \texttt{(bbodyrad} +  $f_a\times$ \texttt{(bbodyrad} + \texttt{compTT))}, where the parameter $f_{\rm a}$ (\texttt{constant} in Xspec) evaluated the enhancement of the persistent emissions during X-ray burst. We subtracted the instrumental background. The persistent spectral shape was assumed unchanged during bursts. Therefore, the parameters of ${\tt bbodyrad + compTT}$ were fixed to the best-fit values listed in Table~\ref{table:preburst}. The absorption column density, $N_{\rm H}$, was fixed as the best-fitting value from the persistent spectra (see Table~\ref{table:preburst}). We only allowed the parameters of \texttt{bbodyrad} and $f_{\rm a}$ to change freely during the fitting. The fitted parameters are shown as red dot in Figs. \ref{Fig:fa_re} and \ref{Fig:fa_re_all}. The $\chi^2$ values are around 1.0, which is displayed in the bottom panel of this figures implying the significant improvements by the $f_{\rm a}$ model compared with the blackbody model, shown as the gray dashed line. 

We note that \citet{Keek2018} also reported the analysis of burst \#1 by using the $f_{\rm a}$ model. They extracted spectra with exposures of 0.03 s at the peak count rate and obtained the maximum radius of 190 km. Compared with their results, we obtained a smaller radius at $154\pm{10}$ km by using spectra with 0.1 s, and smaller blackbody fluence, $2.5\times10^{-7}~ \mathrm{erg~cm^{-2}}$ versus $2.8\times10^{-7}~ \mathrm{erg~cm^{-2}}$. However, the mean $f_{\rm a}$ of PRE phase and maximum temperature in our work are consistent with \citet{Keek2018}.

From the time-resolved spectroscopy, all bursts showed the characteristics of PRE bursts \citep{Galloway2008}. The blackbody radii expanded and reached the maximum within 1 s, meanwhile, the temperature dropped to the minimum. The strong expansion period lasted only $\sim 0.6~ \rm{s}$, and then the blackbody radii began to shrink. When the radii were close to 50 km, the subsequent decrease slowed down, and entered into a plateau of moderate expansion. Finally, when the atmosphere moved back to the NS surface, the blackbody temperature increased to its peak and the blackbody radii decreased to $\sim10$ km, which corresponds to the touchdown moment (i.e., the photosphere moved back to the NS surface). Thirteen bursts exhibited strong radius expansion in excess of a factor 10. The photospheric radius of burst \#8  increased by a factor $>$ 100, that is $r_{\rm{ph}} \approx 1318 ~\rm{km}$, accompanied by the minimum temperature $T_{\rm bb} \approx 0.17 ~\rm keV$, which can be referred to as a superexpansion burst \citep{int2010}. After the onset of the burst the  $f_{\rm a}$ increased rapidly and reached a mean value of $f_{\rm a} \sim 8.5$; then it began to decrease during the moderate expansion phase and returned to unity in the cooling tail. The fitted peak flux, the maximum radius, and the touchdown flux are listed in Table~\ref{table:burst_p}. In addition, the distribution of $f_{\rm a}$ during the PRE phase and  the maximum $f_{\rm a}$ for each burst are shown in Fig.~\ref{Fig:fa_factor}. The maximum $f_{\rm a}$ is $\sim$16 from burst \#7.   We note that the maximum $f_{\rm a}$ for each burst is inversely proportional to the flux persistent emission (see Fig.~\ref{Fig:fa_factor}).

\subsubsection{The disk reflection model}
\label{Sec:spec_diskref}

The relativistic reflection model, \texttt{relxillNS}  \citep{Ludlam19,Connors20}\footnote{\url{https://www.sternwarte.uni-erlangen.de/~dauser/research/relxill/}}, describes the radiation reprocessed from an accretion disk around the NS, in which the seed photons are characterized by a blackbody spectrum. The assumptions of this model are consistent with the  accretion disk illuminated by the burst radiation, which has been successfully explained by the deviations of the burst spectra in 4U 1636--536 and 4U 1730--22 \citep{Zhao22,Lu2023A&A}. In previous works the reflection model and the $f_{\rm a}$ model improved the burst spectral fitting equally well; however, the burst temperatures and bolometric fluxes from the disk reflection model were less than the values from the $f_{\rm a}$ model by the factors of $\sim0.25$ and 2.3, respectively. Since the distances to 4U 1636--536 and 4U 1730--22  were only estimated by using the PRE bursts, it is difficult to distinguish these two models by comparing the measured peak flux to the Eddington flux. In this work, for the first attempt, which is the same method proposed in \citet{Zhao22} and \citet{Lu2023A&A}, we fitted the burst spectra by using the model \texttt{TBabs} $\times$ \texttt{(bbodyrad + relxillNS)}. The persistent emission was regarded as background and was subtracted. The temperature of the seed photons was tied with the blackbody from the burst. The reflection fraction was fixed at -1, which means that all detected emission was attributed to reflection. Assuming a canonical NS mass $M_{\rm NS}=1.4M_{\odot}$ and radius $R_{\rm NS}=10$ km, as well as the derived spin frequency of 275 Hz from quasi-periodic oscillations \citep{White_1997}, we obtained the dimensionless spin parameter, $a=0.13$.  We also tried different spin parameters, $a=0$ and $a=0.3$, which provided consistent results. Therefore, the value of $a$ is not important here. Other parameters were fixed at their default values. Only the normalization of the \texttt{relxillNS} was allowed to change. We found that the mean peak flux is $3.7\times10^{-8}~{\rm erg~s^{-1}~cm^{-2}}$, less than the flux observed with \rxte/PCA from 4U 1820--30, $F_{\rm Edd}\approx5.4\times10^{-8}~{\rm erg~s^{-1}~cm^{-2}}$ \citep{Galloway08,Suleimanov2017}, and also smaller than the Eddington flux, 4.0--6.8 $\times10^{-8}~{\rm erg~s^{-1}~cm^{-2}}$, for a NS with a mass of 1.4--1.8 $M_\odot$ and radius of 10--12 km at a distance of 8.4 kpc.  Subsequently, we left the logarithmic value of the accretion disk density (in units of $\rm{cm^{-3}}$), $\log n$, and the iron abundance of the material in the accretion disk in units of solar abundance, $A_{\rm Fe}$, of the \texttt{relxillNS} model,  free to vary. We found that the solar iron abundance was around $5.0 \pm$1.5 for most bursts, and therefore it was fixed at 5.0.

\begin{figure*}
    
    \centering
        \includegraphics[width=\hsize]{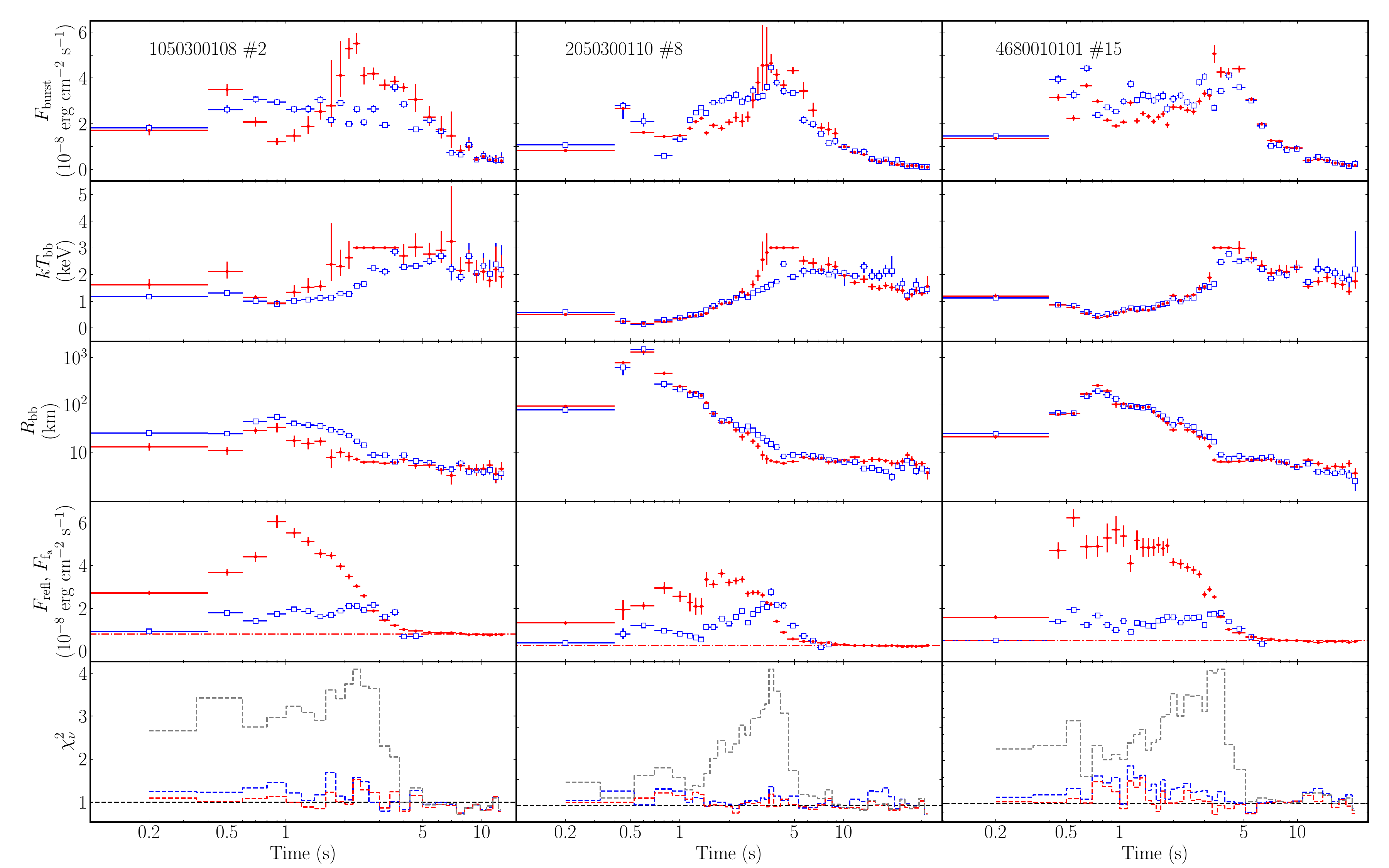}
                \caption{Time-resolved spectroscopy using the $f_{\rm a}$ model (red dot) and the disk reflection model (blue square) for bursts \#2 ({\it left column}), \#8 ({\it middle column}), and \#15 ({\it right column}). Shown in each column, from top to bottom, are the burst bolometric flux, $F_\mathrm{ burst}$; the blackbody temperature, $kT_\mathrm{bb}$, the blackbody radii, $R_\mathrm{bb}$, which were calculated using a distance of 8.4 kpc; the enhanced persistent emission flux and disk reflection flux,  $F_{f_{\rm a}}$ and $F_{\rm{refl}}$, respectively; and the goodness of fit per degree of freedom, $\chi^2_\nu$. The red dashed-dotted lines represent the persistent emission level. The $\chi^2_\nu$ of the blackbody model are plotted as gray dashed line for comparison.} 
        \label{Fig:fa_re}%
\end{figure*}

\begin{figure}
\centering
\includegraphics[width=4cm]{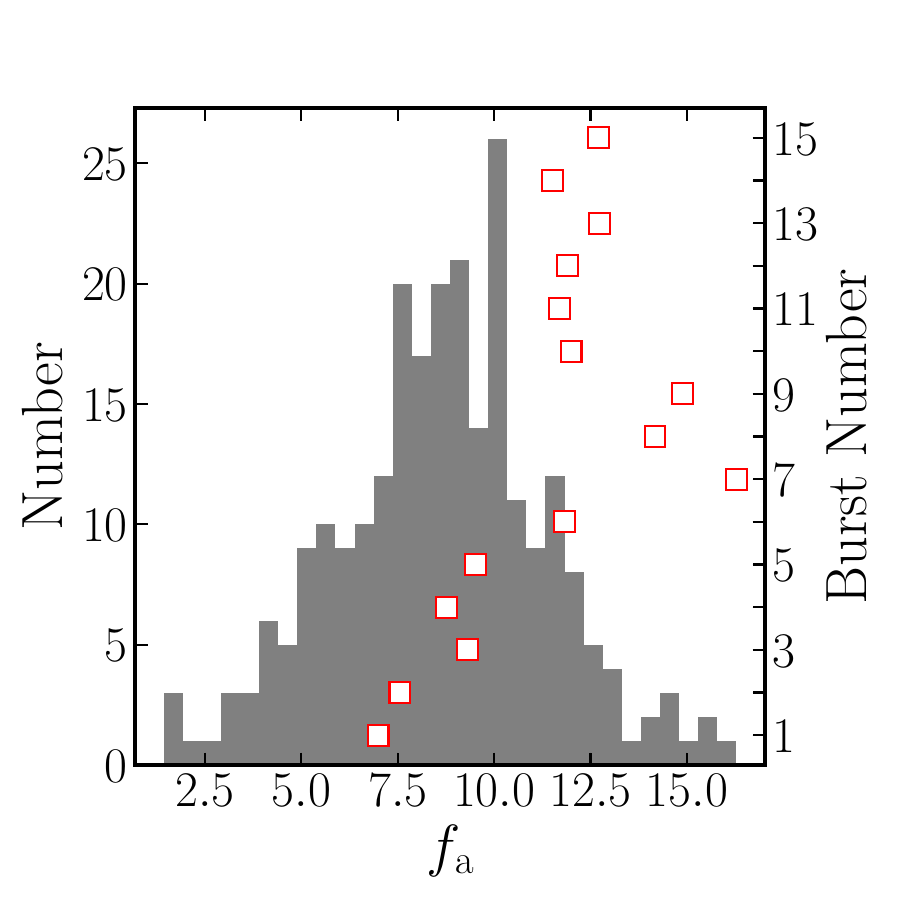}
\includegraphics[width=4cm]{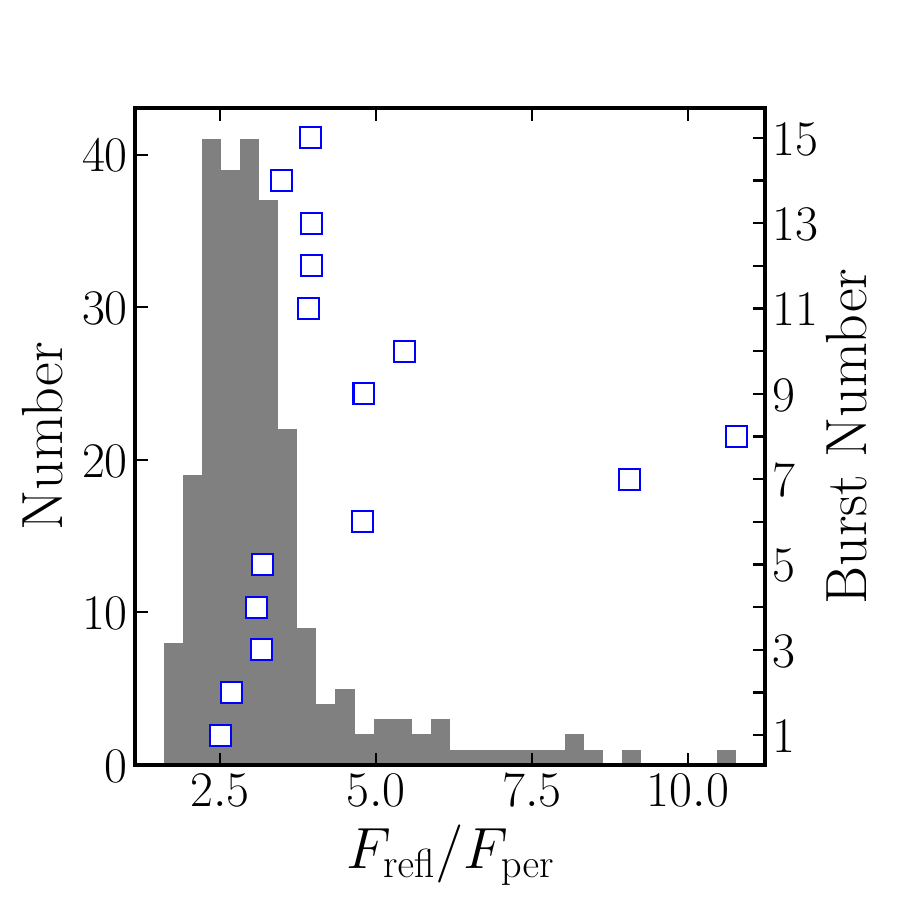}
\includegraphics[width=4cm]{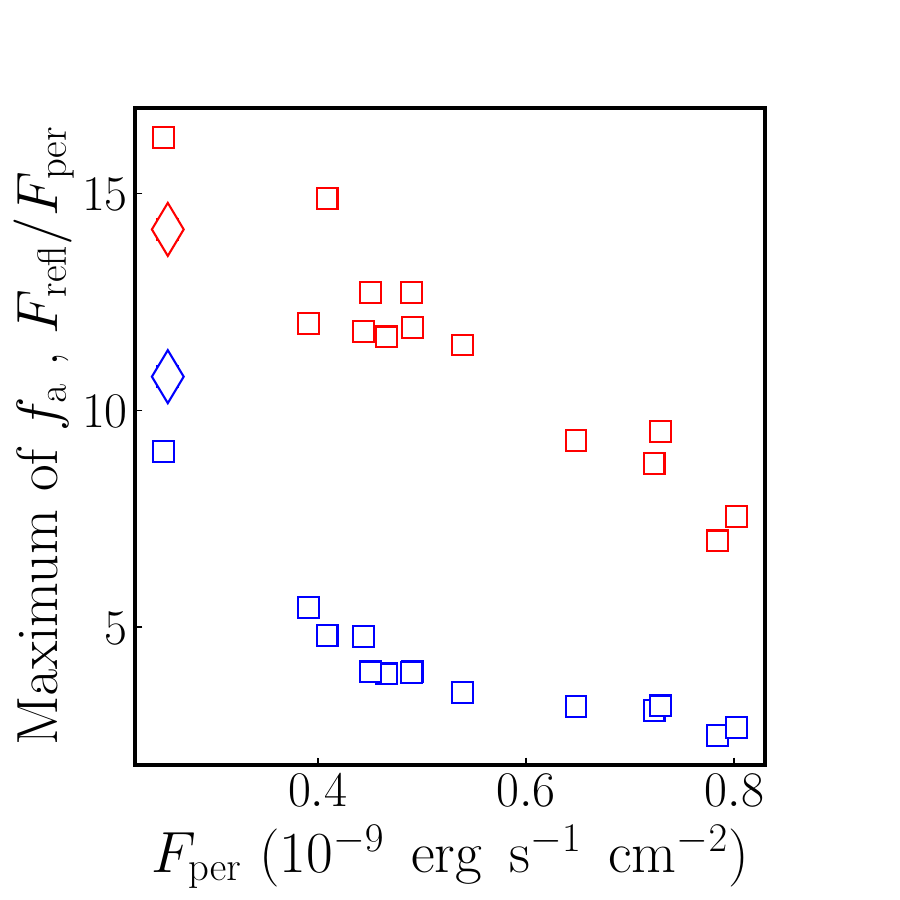}

\caption{Distribution of the $f_{\rm a}$ factor (\textit{top left}) and the reflection fraction (\textit{top right}) during the PRE phase. The right axis shows the maximum $f_{\rm a}$ factor and reflection fraction with square for each burst. In the {\it bottom panel}, maximum $f_{\rm a}$ factor and maximum reflection fraction vs. fluxes of persistent emission. The superexpansion burst is plotted as a diamond. }

      \label{Fig:fa_factor}%
\end{figure}

\begin{figure*}
    
    \centering
        \includegraphics[width=\hsize]{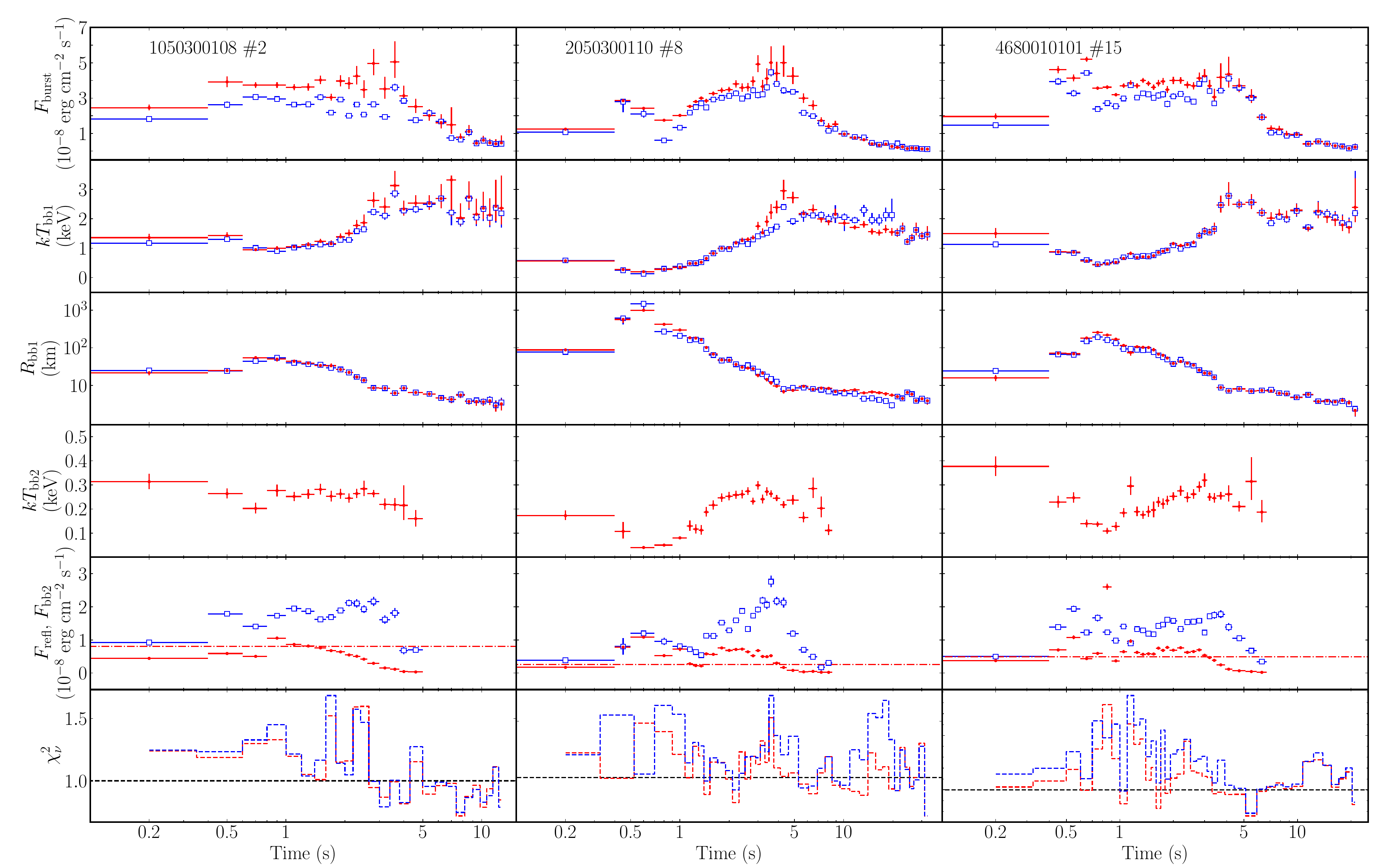}
        \caption{Time-resolved spectroscopy using the double-blackbody model (red dot) and the disk reflection model (blue square) for bursts \#2 ({\it left column}), \#8 ({\it middle column}), and \#15 ({\it right column}), respectively. Show in each column, from top to bottom, are the burst bolometric flux, $F_\mathrm{ burst}$; the hotter blackbody temperature, $kT_\mathrm{bb1}$, the hotter blackbody radii, $R_\mathrm {bb1}$, which were calculated using a distance of 8.4 kpc; the cold blackbody temperature, $kT_{\mathrm {bb2}}$; the cold blackbody flux and disk reflection flux, $F_\mathrm {bb2}$ and $F_{\mathrm{ refl}}$, respectively; and the goodness of fit per degree of freedom, $\chi^2_\nu$.  The red dash-dotted lines represent the persistent emission level. }
        \label{Fig:2bb}%
\end{figure*}

\begin{figure}

    \centering
        \includegraphics[width=\hsize]{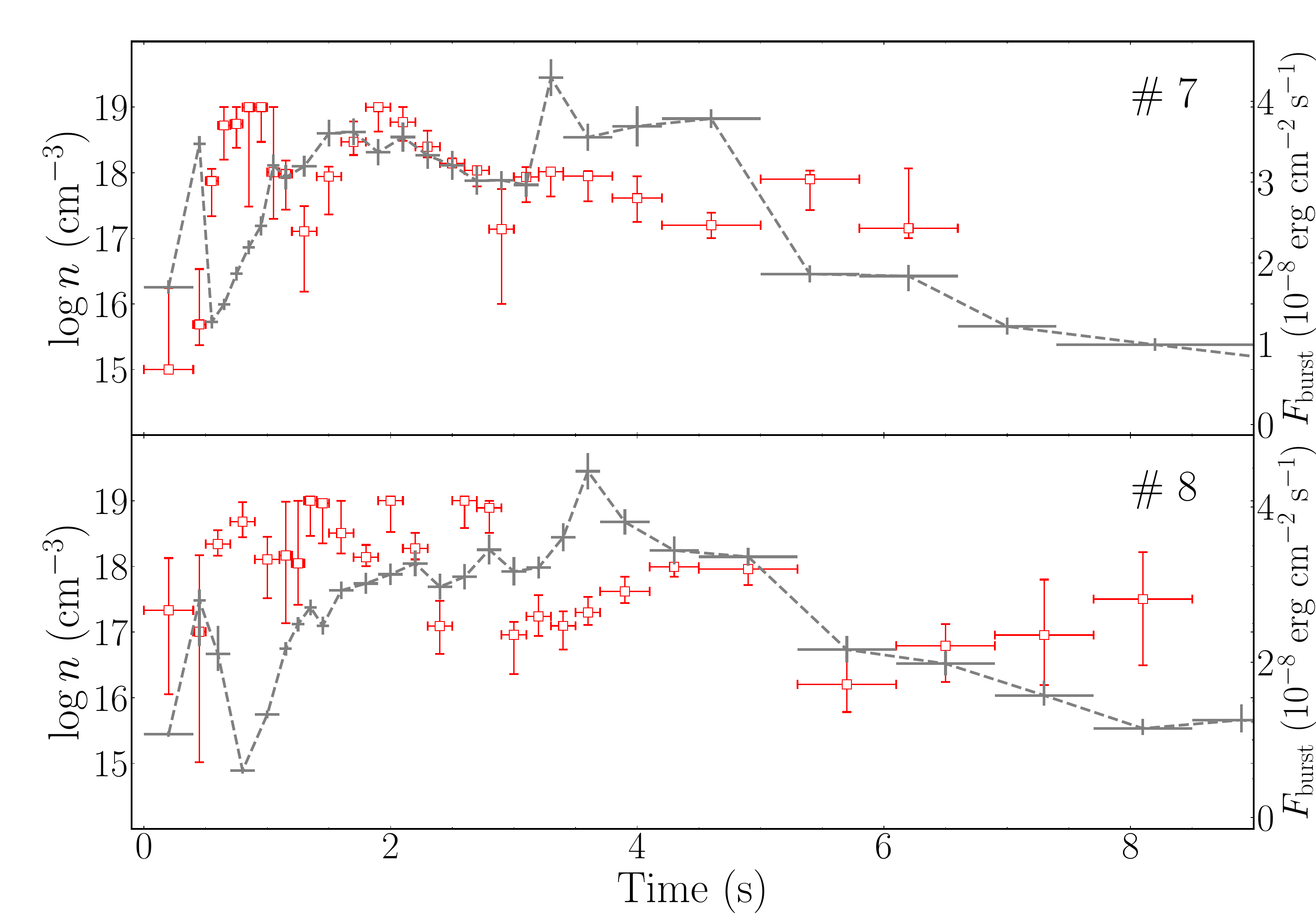}

\caption{Logarithmic value of the accretion disk density obtained from the disk reflection model for bursts \#7 ({\it top}) and \#8 ({\it bottom}).}
      \label{Fig:disk}
\end{figure}

This model yielded a reasonable description of all burst spectra, and the $\chi^2_\nu$ are around 1.0, which means the disk reflection can also explain the derivations properly. In Figs.~\ref{Fig:fa_re}, \ref{Fig:2bb}, \ref{Fig:fa_re_all}, and \ref{Fig:2bb_all}, we compared the fitted parameters of the disk reflection model, which are shown as blue squares, with the enhanced persistent emission model and the double-blackbody model (see Sect.~\ref{Sec:spec_2bb}). The bolometric flux of the \texttt{relxillNS} component, $F_\mathrm{{refl}}$, was calculated in the energy range of 0.1--250 keV. We only showed the $F_\mathrm{{refl}}$ for the first $\sim5$ s for all bursts because they are statistically insignificant for the  reflection component during the cooling tail. Based on the time-resolved spectra, we found that the disk reflection model provides similar parameters from X-ray bursts compared with another two models (see Table~\ref{table:burst_p}). As we found in the $f_{\rm a}$ model, 13 bursts exhibited strong radius expansion in excess of a factor of $>10$, and the superexpansion burst radius reached around $1484\pm{105} ~\rm{km}$. The distribution of the ratio $F_\mathrm{{refl}}/F_\mathrm{{per}}$, and its maximum of each burst  during the PRE phase are displayed in Fig. \ref{Fig:fa_factor}.  We show the $\log n$ evolution of bursts \#7 and \#8 in Fig. \ref{Fig:disk}.  We find that the accretion disk density changes during the burst. For burst \#7, its flux shows two equivalent  peaks at $\sim3.5\times10^{-8}~{\rm erg~cm^{-2}~s^{-1}} $ in 0.5 and 1.5 s after the onset. As a response,  the accretion disk density increases from $10^{15}~{\rm cm^{-3}}$ to the upper limit of the {\tt relxillNS} model, $10^{19}~{\rm cm^{-3}}$,  1 s after the onset, and then it drops to a local minimum of $10^{17}~{\rm cm^{-3}}$ in 0.5 s. In the next 1 s, the accretion disk density reachs $10^{19}~{\rm cm^{-3}}$ again, and decreases to $10^{17}~{\rm cm^{-3}}$ in the cooling tail. The changed in disk density is delayed 0.5 s relative to the burst flux variation. Burst \#8 and other bursts have similar trends but with larger fluctuations.

\subsubsection{Double-blackbody model}
\label{Sec:spec_2bb}
We also employed the double-blackbody model, \texttt{TBabs} $\times$ \texttt{(bbodyrad} + \texttt{bbodyrad)}, to fit the  time-resolved burst spectra. The hotter blackbody  emitted from a neutron star photosphere and the colder one maintained a stable flux through the burst and could possibly arise from the accretion disk. As in the $f_{\rm a}$ model, we found that all bursts show significant PRE, and there are 14 bursts with  $r_{\rm ph}>100$ km. The superexpansion burst of burst \# 8, approached the peak of blackbody radii around $ 1012\pm{53} ~\rm{km}$ with the minimum of the temperature $\sim ~0.17 ~\rm keV$. The colder blackbody component was statistically significant for the first $\sim5$ s for all bursts, and can be neglected during the cooling tail. The parameters fitted by the double-blackbody model are shown as red dots in Fig. \ref{Fig:2bb}. The $\chi^2_\nu$ are displayed in the bottom panels. For each burst the photosphere reached its peak within 1 s. The bolometric fluxes at the peak, $F_{\rm peak}$, and at the touchdown moment, $F_{\rm TD}$, for all burst are listed in Table~\ref{table:burst_p}. The double-blackbody model provides quite similar results of the burst temperature and radius compared with the disk reflection models. This is not surprising because the shape of the {\tt relxillNS} component for the fitted parameters in 0.5--10 keV are close to blackbody.  Since the double-blackbody model is phenomenologically simpler, in the following we focus on this model to analyze the burst properties and the spectral features. We also attempted to include the possible absorption edges around 6--10 keV in the model. We found the edge cannot improve the fit significantly and obtained the absorption depth $\tau_{\rm max} < 3\times10^{-3}$.

We define the duration of the superexpansion or strong photospheric expansion phase, $t_{\mathrm{se}}$, as the time between the first peak of hard light curve and the start of the moderate expansion from the result of time-resolved spectroscopy. The duration of the moderate expansion phase, $t_{\mathrm{me}}$, is calculated as the time between start of the moderate expansion and the time of touchdown. We obtained the burst fluence, $E_{\rm b},$ by summing the measured flux over the burst, and calculated the decay time, $\tau=E_{\rm b}/F_{\rm peak}$. For burst \#10, part of data in the tail were unavailable, therefore, we obtained its burst fluence from two parts, the rise part from the above-mentioned way, and the decay part by fitting the exponential decay of burst flux with the function $F=A\exp{-t/\tau}$,  which is $A\tau$.  The value of $t_{\mathrm{se}}$, $t_{\mathrm{me}}$, $E_{\rm b}$, and $\tau$ are listed in Table~\ref{table:gap}. In all bursts,  $t_{\mathrm{se}}$ lasts $\sim$ 1 s, but all $t_{\mathrm{me}}$ are longer than $t_{\mathrm{se}}$.

\begin{table*}[h!]
\begin{center} 
\caption{Fitted parameters during the peak of bursts by different models.  \label{table:burst_p}}

\resizebox{\linewidth}{!}{\begin{tabular}{ccccccccccccccccc} 
&  \multicolumn{4}{c}{ double-blackbody model}& & \multicolumn{4}{c}{$f_a$ model }& &\multicolumn{4}{c}{ Disk Reflection Model}& \\
\cline{2-5} \cline{7-10} \cline{12-15}\\
{\centering  Burst  } &

{\centering  $F_{\rm peak}^{\rm a}$  }&
{\centering  $kT_{\rm PRE}^{\rm b}$ } &
{\centering  $r_{\rm ph}^{\rm c}$ } &
{\centering  $F_{\rm TD}^{\rm d}$ } &
&
{\centering  $F_{\rm peak}^{\rm a}$  }&
{\centering  $kT_{\rm PRE}^{\rm b}$ } &
{\centering  $r_{\rm ph}^{\rm c}$ } &
{\centering  $F_{\rm TD}^{\rm d}$ } &
&
{\centering  $F_{\rm peak}^{\rm a}$  }&
{\centering  $kT_{\rm PRE}^{\rm b}$ } &
{\centering  $r_{\rm ph}^{\rm c}$ } &
{\centering  $F_{\rm TD}^{\rm d}$ } &
&
\\
 $\#$  &  &$\mathrm{(keV)}$&(km)& && &$\mathrm{(keV)}$&(km)&& &&$\mathrm{(keV)}$&(km)&\\ [0.01cm] \hline
1 &$5.2\pm{1.4}$&$0.59\pm{0.02}$&$151\pm{9}$  &$4.1\pm{0.7}$&&$5.5\pm{1.4}$&$0.45\pm{0.02}$&$154\pm{10}$ &$4.5\pm{1.4}$&&$3.9\pm{0.1}$&$0.62\pm{0.01}$&$126\pm{10}$ &$3.9\pm{0.2}$\\
2 &$5.1\pm{1.2}$&$0.95\pm{0.04}$&$55\pm{5}$   &$4.9\pm{0.7}$&&$5.5\pm{0.5}$&$0.93\pm{0.11}$&$33\pm{8}$   &$4.2\pm{0.3}$&&$3.6\pm{0.2}$&$0.90\pm{0.03}$&$55\pm{4}$   &$3.6\pm{0.2}$\\
3 &$4.7\pm{0.3}$&$0.51\pm{0.02}$&$192\pm{14}$ &$3.4\pm{0.6}$&&$5.7\pm{0.5}$&$0.38\pm{0.02}$&$215\pm{25}$ &$4.4\pm{0.3}$&&$4.2\pm{0.2}$&$0.53\pm{0.02}$&$157\pm{14}$ &$3.4\pm{0.2}$\\
4 &$5.0\pm{0.6}$&$0.59\pm{0.02}$&$142\pm{9}$  &$3.3\pm{0.5}$&&$6.1\pm{0.4}$&$0.47\pm{0.02}$&$143\pm{12}$ &$4.1\pm{0.3}$&&$3.9\pm{0.1}$&$0.63\pm{0.02}$&$116\pm{7}$  &$3.3\pm{0.1}$\\
5 &$4.9\pm{1.2}$&$0.68\pm{0.02}$&$108\pm{8}$  &$4.8\pm{0.9}$&&$5.0\pm{0.4}$&$0.60\pm{0.07}$&$72\pm{10}$  &$4.0\pm{0.4}$&&$3.5\pm{0.2}$&$0.70\pm{0.02}$&$92\pm{6}$   &$3.5\pm{0.2}$\\
6 &$5.8\pm{0.3}$&$0.42\pm{0.01}$&$284\pm{16}$ &$5.8\pm{0.3}$&&$6.1\pm{0.3}$&$0.37\pm{0.01}$&$284\pm{21}$ &$6.1\pm{0.3}$&&$5.0\pm{0.2}$&$0.42\pm{0.01}$&$271\pm{15}$ &$5.0\pm{0.2}$\\
7 &$5.3\pm{1.3}$&$0.35\pm{0.02}$&$340\pm{23}$ &$5.3\pm{1.0}$&&$5.9\pm{0.4}$&$0.27\pm{0.01}$&$397\pm{36}$ &$4.7\pm{0.4}$&&$4.3\pm{0.2}$&$0.33\pm{0.01}$&$331\pm{22}$ &$4.3\pm{0.2}$\\
8 &$5.0\pm{0.9}$&$0.20\pm{0.01}$&$1012\pm{53}$&$5.0\pm{0.8}$&&$4.6\pm{0.4}$&$0.17\pm{0.01}$&$1318\pm{95}$&$4.5\pm{0.2}$&&$4.5\pm{0.2}$&$0.13\pm{0.02}$&$1484\pm{105}$&$4.5\pm{0.2}$\\
9 &$4.7\pm{1.0}$&$0.49\pm{0.02}$&$195\pm{19}$ &$4.7\pm{0.8}$&&$3.6\pm{0.1}$&$0.39\pm{0.01}$&$250\pm{18}$ &$3.5\pm{0.3}$&&$3.7\pm{0.1}$&$0.50\pm{0.02}$&$166\pm{20}$ &$3.7\pm{0.1}$\\
10&$5.0\pm{0.7}$&$0.40\pm{0.01}$&$317\pm{17}$ &$4.4\pm{0.6}$&&$5.3\pm{0.3}$&$0.36\pm{0.01}$&$309\pm{24}$ &$4.4\pm{0.2}$&&$4.1\pm{0.2}$&$0.40\pm{0.02}$&$301\pm{19}$ &$4.1\pm{0.2}$\\
11&$4.9\pm{0.9}$&$0.45\pm{0.01}$&$252\pm{13}$ &$4.9\pm{0.8}$&&$4.7\pm{0.8}$&$0.40\pm{0.01}$&$253\pm{13}$ &$4.7\pm{0.8}$&&$4.3\pm{0.1}$&$0.48\pm{0.01}$&$226\pm{13}$ &$4.3\pm{0.1}$\\
12&$5.4\pm{1.2}$&$0.49\pm{0.01}$&$219\pm{12}$ &$5.4\pm{0.9}$&&$5.3\pm{1.1}$&$0.45\pm{0.02}$&$166\pm{23}$ &$5.0\pm{0.3}$&&$5.0\pm{0.2}$&$0.49\pm{0.02}$&$103\pm{11}$ &$5.0\pm{0.2}$\\
13&$4.8\pm{0.3}$&$0.52\pm{0.02}$&$201\pm{12}$ &$3.6\pm{0.5}$&&$4.6\pm{0.2}$&$0.39\pm{0.02}$&$198\pm{15}$ &$4.2\pm{0.3}$&&$4.5\pm{0.2}$&$0.54\pm{0.02}$&$175\pm{13}$ &$3.9\pm{0.1}$\\
14&$4.9\pm{0.3}$&$0.44\pm{0.01}$&$243\pm{13}$ &$4.8\pm{0.9}$&&$4.7\pm{0.3}$&$0.40\pm{0.01}$&$256\pm{17}$ &$4.3\pm{0.3}$&&$4.4\pm{0.1}$&$0.46\pm{0.01}$&$238\pm{11}$ &$4.0\pm{0.1}$\\
15&$5.2\pm{0.2}$&$0.43\pm{0.01}$&$256\pm{14}$ &$4.4\pm{0.8}$&&$5.1\pm{0.4}$&$0.40\pm{0.01}$&$255\pm{18}$ &$4.3\pm{0.3}$&&$4.4\pm{0.1}$&$0.45\pm{0.02}$&$193\pm{19}$ &$4.2\pm{0.1}$\\
\hline  
\end{tabular} }

\end{center}

\tablefoot{
\tablefoottext{a}{The bolometric peak flux of each burst in units of $10^{-8}~ \mathrm{erg~s^{-1} cm^{-2}}$. }\\
\tablefoottext{b}{ The minimum blackbody temperature of each burst.}\\
\tablefoottext{c}{The maximum blackbody radii.}\\
\tablefoottext{d}{ The bolometric touchdown flux in units of $10^{-8}~ \mathrm{erg~s^{-1} cm^{-2}}$.} \\
}
\end{table*}

\section{Discussion}
\label{Sec:discussion}

In this work we analyzed the \nicer\ observations collected during 2017--2023 from 4U 1820--30. We extracted the light curves and produced the HID and CCD of the persistent emissions from \nicer\ observations. The persistent emission can be fitted by a combination of \texttt{bbodyrad} and \texttt{compTT}.
We observed 15 type I X-ray bursts. No significant burst oscillations in these bursts were found. All bursts occurred in the low spectra state, same as the phenomenon found by \rxte\ \citep{Chou2001ApJ,int2012A&A}. They attributed that in the high state the thermonuclear burning is stable due to the high accretion rate, implying the quenching of the unstable burning activity.

\subsection{Burst spectral fitting}
\label{Sec:fitting}

All the time-resolved burst spectra in the first 5 s showed significant deviations from the blackbody model, which can be explained by the $f_{\rm a}$ model, the reflection model, or the double-blackbody model, providing comparable $\chi^2_\nu$ values close to unity. We can distinguish the physically motivated and self-consistent model based on the fitted results. 

\citet{Keek2018} found that there is a strong excess at the lowest energies of the spectra in burst \#1. This excess is  nicely described by enhanced Comptonization emission ($f_{\rm a}$ model) \citep{Keek2018}. They foundthat the Comptonization component in the model becomes six times brighter during the burst, which is substantially more than the model prediction. These authors speculated that it could be caused by the burst emission being reprocessed by the NS environment.  
We also found that the  blackbody model fitted the time-resolved spectra poorly for all 15 X-ray bursts at first $\sim5$ s with $\chi^2_\nu$$ > 2$.  Even though the $f_{\rm a}$ model fit the burst spectra very well, it is not a self-consistent way to explain the derivation from black body (see  discussions in \citet{Keek2018}). We obtained that the mean $f_{\rm a}$ factor of all burst spectra fitted by the $f_{\rm a}$ model is 10, by assuming only the amplitude of the persistent emissions free to change. From the persistent spectra, the persistent fluxes increased by a factor of 6 from the low state to the high state. The spectral parameters (i.e., the temperatures of blackbody and seed photons) also changed. Some hints suggest that the persistent spectral shape have been also varied during bursts \citep{Degenaar16,Koljonen16,Kajava17,Suleimanov2017}. Therefore, some spectral parameters could also change during X-ray bursts if the persistent emission increased several times from pre-burst.  Moreover, the $f_{\rm a}$ factor measured at the dips for bursts \#7 and \#8 are $>10$, which contradicts the observations (i.e., there should be some photons left after the persistent emission subtracted.) 

\begin{figure}
\centering
\includegraphics[width=\hsize]{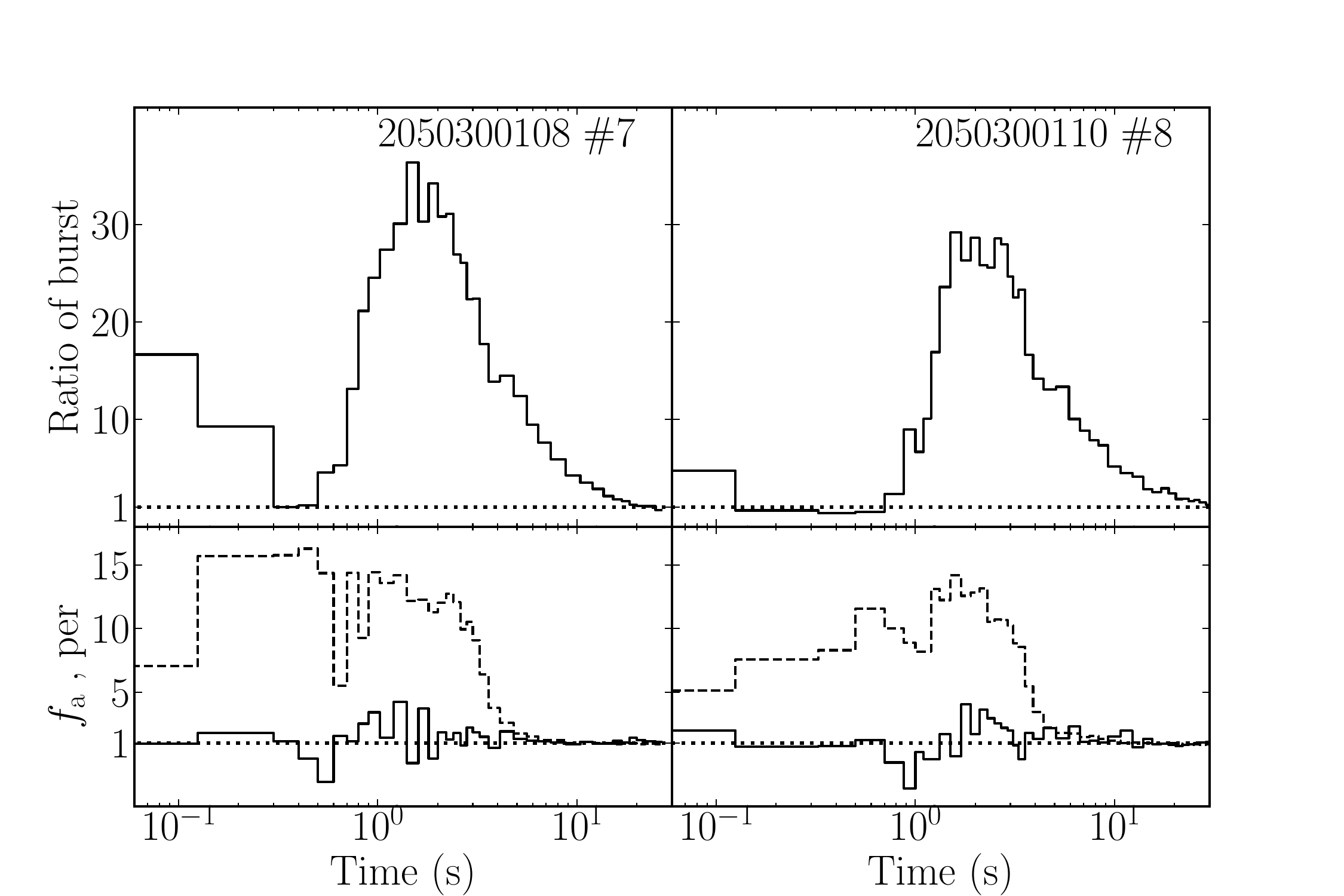}
\caption{ Comparison between simulations and observations for bursts \#7 and \#8. {\it Top}: Ratio of burst counts (solid line) generated by the \texttt{fakeit} command to the persistent emission in 3--10 keV for bursts \#7 (\textit{left}) and \#8 (\textit{right}). %
{\it Bottom}, Ratio between the count rates with the burst contribution subtracted and the pre-burst emissions in 3--10 keV (solid line). The $f_{a}$ is also shown (dashed line) as a comparison. }
      \label{Fig:dip_fak}
\end{figure}

In addition, for each burst we estimated the count rates contributed solely from the burst radiation by using the fitted burst parameters and the ARF and RMF of \nicer. We then reconstructed the burst light curves in 3--10 keV.   For bursts \#7 and \#8, the burst  count rates are normalized to the pre-burst light curves in 3--10 keV (see top panels in Fig.~\ref{Fig:dip_fak}). During the dips, the burst count rates are close to the pre-burst levels, implying that the persistent emissions did not increase in 3--10 keV during the bursts. In the bottom panels in Fig.~\ref{Fig:dip_fak}, we subtracted the burst count rates from the total  burst light curves  in 3--10 keV and normalized to the pre-burst emissions. The dotted lines in the bottom panels in Fig.~\ref{Fig:dip_fak} show that the count rates during the bursts are exactly the same as the pre-burst emissions. For the first 5 s during the bursts, the $f_{\rm a}$ model provides the enhancement of the persistent emission at a factor of 5--15 in 3--10 keV.   However, the count rates of the persistent emission (solid line) during the bursts are far below the $f_{\rm a}$ model prediction (dashed line). The rest of the bursts have the same properties.    This strengthens the idea that the $f_{\rm a}$ model might not be the proper manner to handle the persistent emission that is variable during X-ray bursts. Therefore, we conclude that the $f_{\rm a}$ model cannot explain the deviations properly.

Subsequently, the excesses are fitted well by using the double-blackbody model and  the disk reflection model,  which were adopted previously to fit the X-ray burst spectra from 4U 1820--30  \citep{Keek2018}, SAX J1808.4--3658 \citep{bult2021}, and 4U 1636--536 and 4U 1730--22 \citep{Zhao22,Lu2023A&A}. The double-blackbody and reflection models give a roughly equivalent interpretation that the soft excess tracks an interaction between the burst emission and the neutron star environment. For both models the burst blackbody fluxes are very close because the shape of these two models are quite similar in 0.5--10 keV.  Since the double-blackbody model was proposed phenomenologically, we suggest that only the reflection model provides a physically motivated and self-consistent explanation to the observations.   The reflection component can contribute $\sim$ 30\% of the total burst emission in the strong PRE phase. Moreover, we found that the accretion disk density changed as a response to the burst flux with 0.5 s delay. Compared with the $f_{\rm a}$ model, they provide a comparable goodness of fit, resulting in similar burst blackbody evolutions, but the maximum fluxes obtained from the $f_{\rm a}$ model are slightly higher than the results from the reflection model and the double-blackbody model. The maximum fluxes of the latter two models are slightly lower than the Eddington flux observed with RXTE (PCA) from 4U 1820--30 \citep{Galloway08,Suleimanov2017}, $F_{\rm Edd}\approx5.4\times10^{-8}~{\rm erg~s^{-1}~cm^{-2}}$.

\subsection{superexpansion and strong photospheric expansion}
\label{Sec:spre}

In previous works, most of the superexpansion bursts were captured by Wide Field Cameras \citep[WFCs;][]{Jager97} on BeppoSAX or  the Proportional Counter Array \citep[PCA;][]{Jahoda06} on \rxte\ \citep{int2010}. All these instruments are sensitive to photon energy above 2--3 keV. Around the maximum radius of the superexpansion, the data were usually unavailable due to the emitted blackbody temperature lower than 0.5 keV and most of photons were undetectable.  Thanks to the sensitivity of \nicer\ down to 0.2 keV and relativity low interstellar absorption toward 4U 1820--30, we have the chance to study the photospheric radius evolution across the whole burst. From the time-resolved burst spectroscopy, we found one superexpansion burst (\#8),  $r_{\rm ph}\approx1500$ km, which is also the first superexpansion burst observed by \nicer; 13 strong photospheric expansion bursts, $r_{\rm ph}\approx100-300$ km,  and one moderate PRE burst (\#2), $r_{\rm ph}\approx50$ km. Moreover, we measured the blackbody temperature of 0.2 keV at the largest expansion radius for the superexpansion burst, which is the lowest in all X-ray bursts. This is also the first direct measurement of the full superexpansion stage for the radius above $10^3$ km.   %

Such large differences in photospheric expansion do not often appear in the same source, and we wondered what factors made the difference in the three kinds of expansion. What factors make the difference of three kinds of expansion? \citet{int2012A&A} proposed that the accretion geometry and the accretion rate can strongly influence the burst photospheric expansion. If the accretion rate is reduced or the geometry changes, it is possible to reduce the pressure on the photosphere. The local accretion rate per unit area onto the NS can be calculated from its pre-burst emission \citep{Galloway08} as
\begin{equation}
\begin{split}
    \Dot{m}
    &=\frac{L_\mathrm{ {per}}(1+z)}{ 4\pi R_{\rm NS}^{2}(GM_{\rm NS}/R_{\rm NS})}\\
    &\approx 6.7\times 10^{3}\biggl(\frac{F_\mathrm{{per}}}{10^{-9}\mathrm{~ergs ~ cm^{-2}~s^{-1}}}\biggr)\biggl(\frac{d}{10\mathrm{~ kpc}}\biggr)^{2}\biggl(\frac{M_{\rm NS}}{1.4M_{\odot}}\biggr)^{-1}\\
    &\quad\times\biggl(\frac{1+z}{1.31} \biggr)\biggl(\frac{R\mathrm{_{NS}}}{10\mathrm{~ km}}\biggr)^{-1}\mathrm{g~cm^{-2}}\mathrm{~s^{-1}},
    \label{eq:lo_accration}
\end{split}      
\end{equation}\\ 
where $F_{\rm per}$ is the persistent flux listed in Table~\ref{table:burst_p}.  
Using equation~\ref{eq:lo_accration} and the observed persistent flux, we obtained $\Dot{m}$, which is listed in Table \ref{table:gap}. We assume the local Eddington accretion rate $\Dot{m}_{\rm Edd}=(8.8\times10^{4})[1.7/(X+1)]~\mathrm{g~cm^{-2}~s^{-1}}$. From panels ({\it a}) and ({\it b}) of Fig.~\ref{Fig:Accretion}, we also note that the maximum expansion radii are larger at  harder color (or lower local accretion rate), which means that the superexpansion burst occurred in the more extreme low state.  The superexpansion burst \#8 and the largest strong photospheric expansion burst \#7 had the lowest local accretion rate, $\sim8\%~\Dot{m}_{\rm Edd}$, which is 3.1 times smaller than burst \#2  with  $r_{\rm ph}\sim55$ km. The accretion rates of other strong photospheric expansion bursts are at least two times larger than the superexpansion burst. Changes in the geometry of the inner accretion flow can lead to changes in spectral shape, as measured for instance via the color in Fig.~\ref{fig:HID}. The ram pressure of the accretion disk may be too low to counteract the expansion when the accretion flow geometry near the NS surface is spherical and the accretion rate is low \citep{int2012A&A}.

The luminosity at the base of the envelope, $L_b^\infty$, can strongly affect the expanded photospheric radius \citep{Paczynski1986ApJ}. If $L_b^\infty$ is smaller than the Eddington limit, $L_{\rm Edd}$, it will produce a static expanded atmosphere.  
On the other hand, a strong PRE burst, $L_b^\infty  \gtrsim 1.05L_{\rm Edd}$, could generate a super-Eddington wind, which can cause the matter outflow from the NS surface. Higher $L_b^\infty$ corresponds to larger expanded radius.\citet{Paczynski1986ApJ} proposed a relativistic model of the wind driven by X-ray bursts, and found that all the super-Eddington energy flux is used to gently blow out matter and that the photospheric radius of an outflowing envelope is always at least 100 km. \cite{Yu2018} presented a hydrodynamic simulation of spherically symmetric super-Eddington winds, showing that the photosphere reaches more than 100 km at first second, independent of the burst duration and ignition depth. This is consistent with our results that the photospheric radius reached its maximum  within 1 s after the burst onset and there is no obvious correlation between the photospheric expansion radius and the ignition depth (see  Table~\ref{table:gap}). The duration of the superexpansion phase always lasts a few seconds, independent of the burst duration \citep{int2012A&A}. It was suggested that it corresponds to a transient stage in the wind’s development. A shell of initially opaque material is ejected to large radii by the burst in this stage, and then it becomes optically thin and the observer suddenly sees the underlying photosphere of the already formed steady-state wind. Recently, \citet{Guichandut2021ApJ} calculated steady-state models of radiation-driven super-Eddington winds and found the transition from static expanded envelopes ($r_{\rm ph}\leq 50-70 \rm ~km$ for $L_b^\infty	\lesssim L_{\rm Edd}$) to radiation-driven stellar wind ($r_{\rm ph}\approx 100 - 1000 \rm ~km$ for $L_b^\infty	\gtrsim 1.05L_{\rm Edd}$). Based on \citet{Guichandut2021ApJ}, in our case,  burst \#2 corresponds to a static envelope, and the other bursts  formed steady-state winds.  The unstable burning of pure helium releases nuclear energy of $10^{18}~{\rm erg~g^{-1}}$, which set the upper limit of the mass-loss rate of wind to be $2\times10^{18}~{\rm g~s^{-1}}$, corresponding to $L_b^\infty  \approx 2.5 L_{\rm Edd}$. \citet{Guichandut2021ApJ} found that the maximum photospheric radius that the wind can reach is around $10^3$ km for this luminosity, which is consistent with the superexpansion burst observed in 4U 1820--30.

From the time-resolved spectroscopy, we found that the strong photospheric expansion phase is similar to the superexpansion phase (the $t_{\mathrm{se}}$ lasts $\sim$ 1 s), which is the stage in the wind’s development. Due to the difference in the impact pressure of the accretion flow, the observed expansion radius of the photosphere is different. Only one expansion phase can be seen in burst \#2, which is a reasonable result, because its expansion radius is less than the lowest outflow radius. The photospherical shell cannot escape the gravitational binding, and it appears as an expansion phase in observation. We note that the strong photospheric expansion is similar to the superexpansion, a strong expansion phase followed by a moderate expansion phase. \citet{int2010} proposed that the ejected material shell may be truncated by heavy elements in the wind, which has been observed at a smaller photospheric radius. The heavy elements can produce absorption edges around 6--10 keV, which are difficult to be observed by \nicer\ due to its sensitivity up to $\sim 10$ keV.

\subsection{Dips of the burst light curves}
\label{Sec:dip}

For the light curves with energies in excess of 3 keV, the dips appeared during the peaks of all bursts except for burst \#2 (see Fig.~\ref{fig:lc}). We also found there is a dip in burst \#8 in the 0.5--10 keV light curve after the onsetat $\sim$ 0.5 second, and there is a plateau for bursts \#6, \#7 and \#15 at the same time. Most of the dips can be understood as being due to the spectral softening. The  PRE bursts were so extreme that the peak of the blackbody spectrum moved to lower temperatures, in tandem with the expansion, and out of the X-ray band introducing the appearance of dip in the burst profiles \citep{Tawara1984,zand2011arXiv}. Many of the reported detections of superexpansion PRE bursts to date have been made with data from instruments with a hard bandpass such as \rxte/PCA, which has a minimum sensitive at 3 keV. In previous observations, there was a lack of data during the most extreme phase. \nicer\ is ideal for observing the soft X-ray expansion phases of PRE bursts, whose bandpass extends down to 0.2 keV. Fortunately, the count rates were very high in the bandpass with \nicer\ less than 3 keV. We obtained the minimum temperatures of 0.2--0.7 keV by fitting the time-resolved spectroscopy (see Table~\ref{table:burst_p}). The bursts with higher temperatures have shallower dips in the light curves. The minimum temperature of burst \# 8 is $0.20\pm{0.01}$ keV obtained by the double-blackbody model, which is the lowest observed so far to our knowledge. For burst \#2, the blackbody temperature was 0.93 keV at the maximum photospheric radius, which was the highest of all 15 bursts. Therefore, the dip did not appear distinctly. In other bursts, the blackbody temperatures were in the range of 0.35--0.69 keV.

\begin{figure}
\centering
\includegraphics[width=4cm]{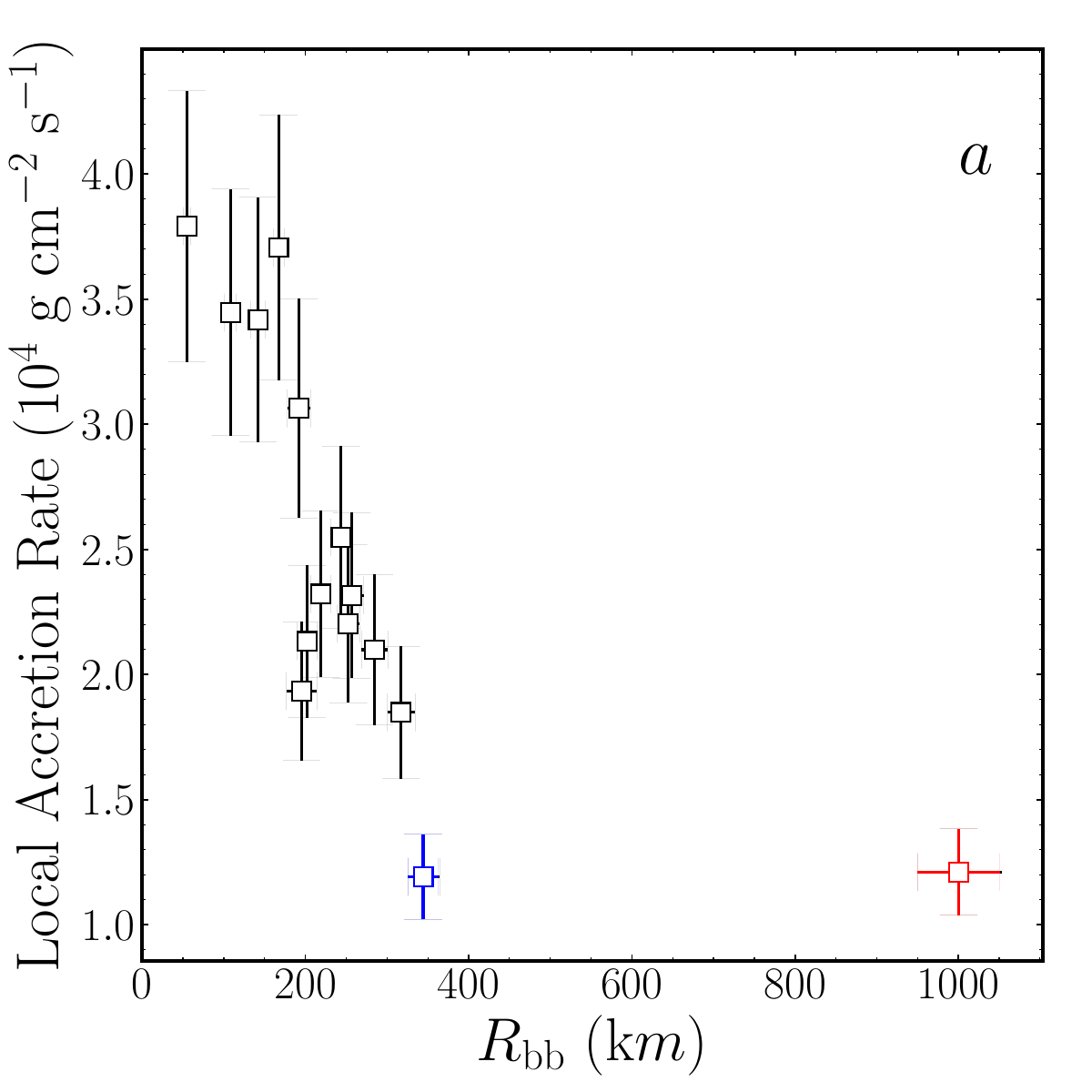}
\includegraphics[width=4cm]{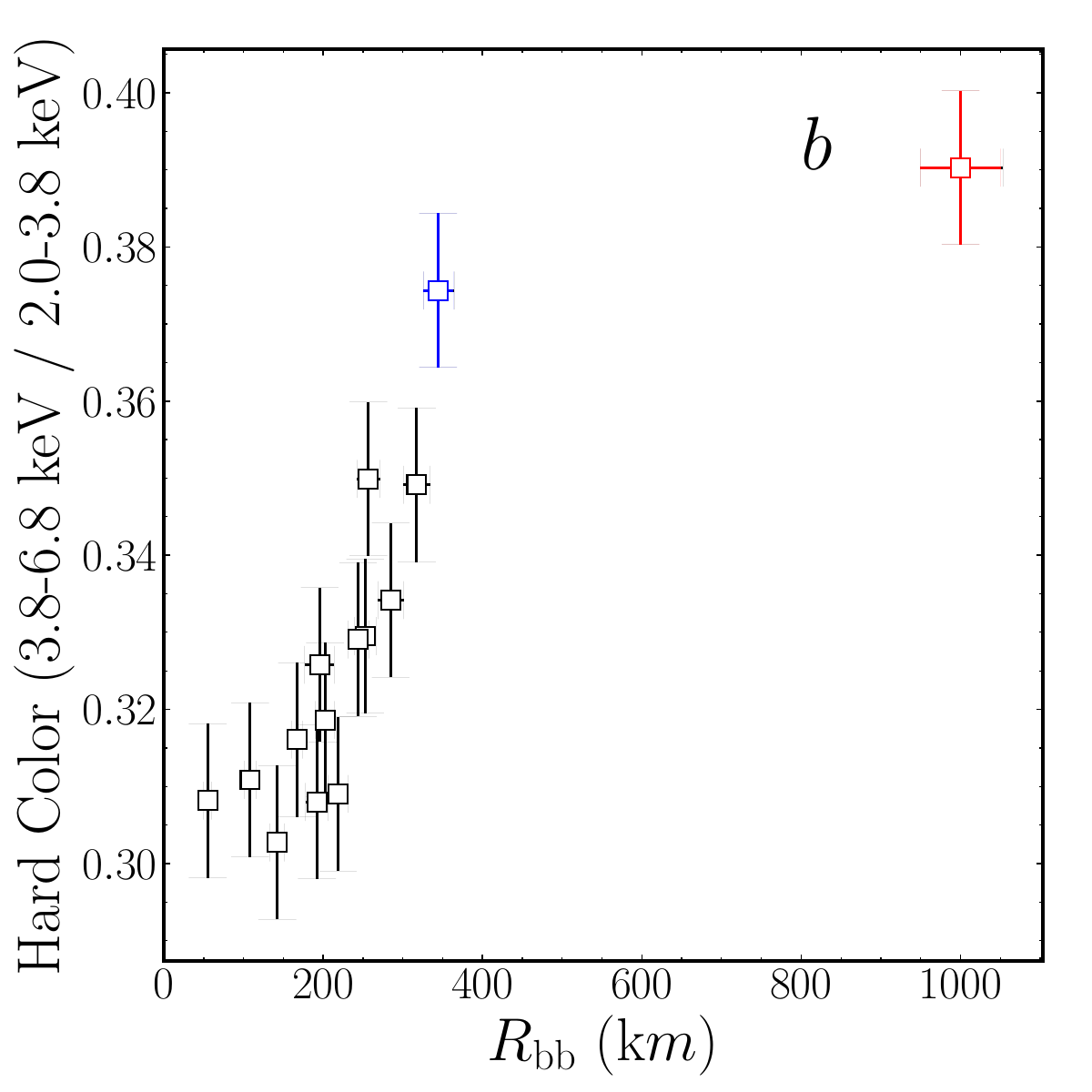}
\includegraphics[width=4cm]{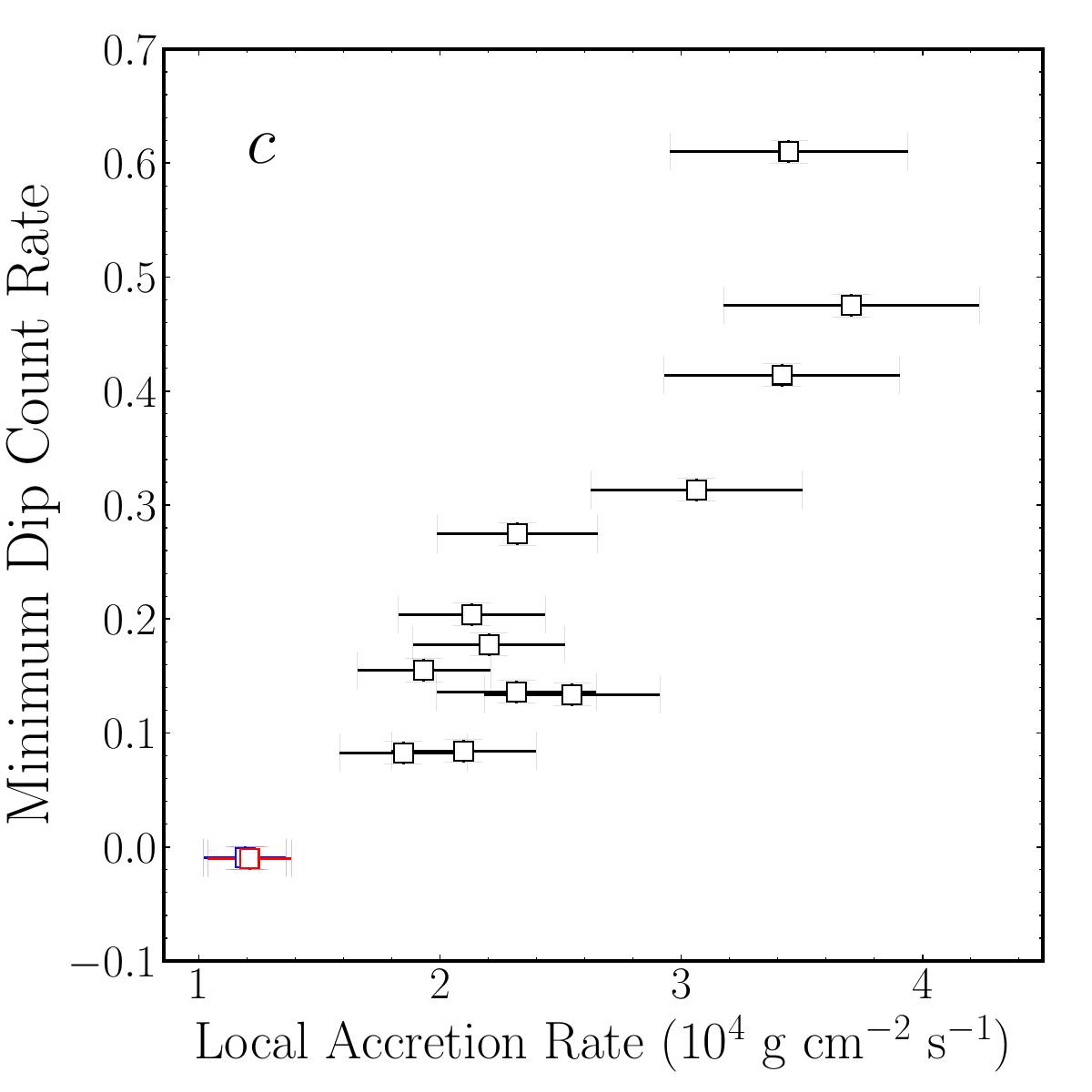}
\includegraphics[width=4cm]{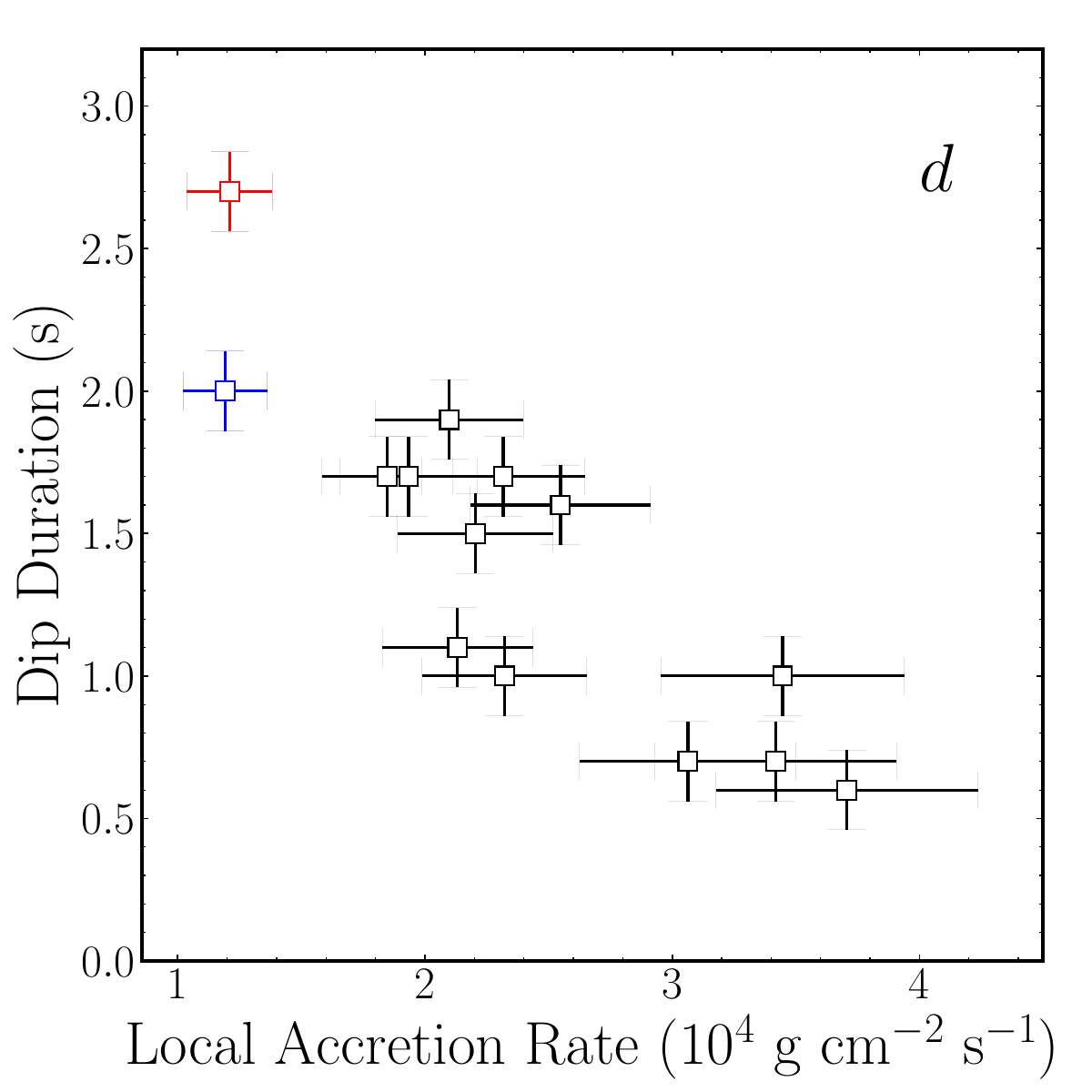}

\caption{Relations of burst properties vs. the persistent emissions, ({\it a}) $\dot{m}-r_{\rm ph}$, ({\it b}) Hard color--$r_{\rm ph}$, ({\it c}) Dip count rate--$\dot{m}$, ({\it d}) Dip duration--$\dot{m}$. Bursts \#7 and \#8 are plotted as a blue and red open square, respectively.}
      \label{Fig:Accretion}
\end{figure}

Spectral softening cannot fully account for the count rates that declined to below the pre-burst level in bursts \#7 and \#8.  A plausible physical interpretation is that the accretion radiation is cut short by the expanding shell or that the shell engulfs the accretion flow so that it is temporarily blocked from our view \citep{int2010,int2012A&A}. The strong PRE bursts may cause the disk to distort itself through a warping instability, which obscures the inner disk \citep{Ballantyne2023}. 
In Fig.~\ref{Fig:Accretion} we find that the minimum count rate during the dip is positively correlated with $\Dot{m}$.  For bursts \#7 and \#8, $\Dot{m}\sim8\%~\Dot{m}_{\rm Edd}$  are the lowest accretion rates,  indicating that accretion flows are more susceptible when disturbed by the burst.

Alternatively, the super-Eddington flux can disrupt the inner disk and stop the accretion temporarily. The disruption of the accretion disk by the ejected shell has been detected in some superexpansion bursts \citep{Strohmayer_2002,int2011A&A,int2019A&A}. \nicer\ also observed the light curve  drop to below the persistent emission in an intermediate-duration burst tail of IGR J17062--6143 \citep{bult2021}.  This dip was interpreted as being due to the accretion disk and the corona being destroyed by the energetic X-ray burst.  \citet{int2012A&A} used the order-of-magnitude calculation to show that the shell may have enough momentum to sweep the inner 100 km of the accretion disk, when the column thickness of the shell at launch is around $10^{7} \mathrm{~g~ cm^{-2}}$.

\subsection{Comparison of recurrence time and $E_{\rm b}-F_{\rm per}$ relation with theoretical models}

The observed recurrence times are listed in Table~\ref{table:gap}, which are only upper limits due to data gaps between bursts. The observed recurrence time between bursts \#1 and \#2, $\Delta T_{\rm rec}=2.2$ hr, and bursts \#3 and \#4, $\Delta T_{\rm rec}=2.8$ hr, are comparable with previous works. Therefore, we consider them as two pairs of successive bursts even though data gaps exists. We compare the observed and predicted recurrence time.  The burst ignition depth is estimated via 
\begin{equation}
y_{\rm ign}=\frac{4\pi E_\mathrm{{b}}d^{2}(1+z)}{4\pi R_{\rm NS}^{2}Q\mathrm{_{nuc}}}, \label{eq:ign}
\end{equation} 
where the nuclear energy generated for pure helium is $Q_\mathrm{{nuc}} \approx 1.31 \mathrm{~MeV~nucleon^{-1}}$ for $X=0$  \citep{Goodwin19}. 
Once the ignition depth is known, we calculate the recurrence time between bursts by using the equation $\Delta t_{\mathrm{rec}}=(y_{\mathrm{ign}} / \Dot{m})(1+z)$. The predicted recurrence times are $1.3\pm0.5$ and $1.5\pm0.5$ hr, which are smaller than the observed recurrence times of 2.2 and 2.8 hr, respectively.  If the factor, $\xi_p^{-1}\sim1.5$, is considered to account for the anisotropy of X-ray emission, the predicted recurrence time can match the observed values  \citep{He16}. We also note that the recurrence time between bursts \#4 and \#5 is 6.3 hr. One or two bursts could be missed by comparing it with the predicted recurrence time.

\begin{figure}
\includegraphics[width=\hsize]{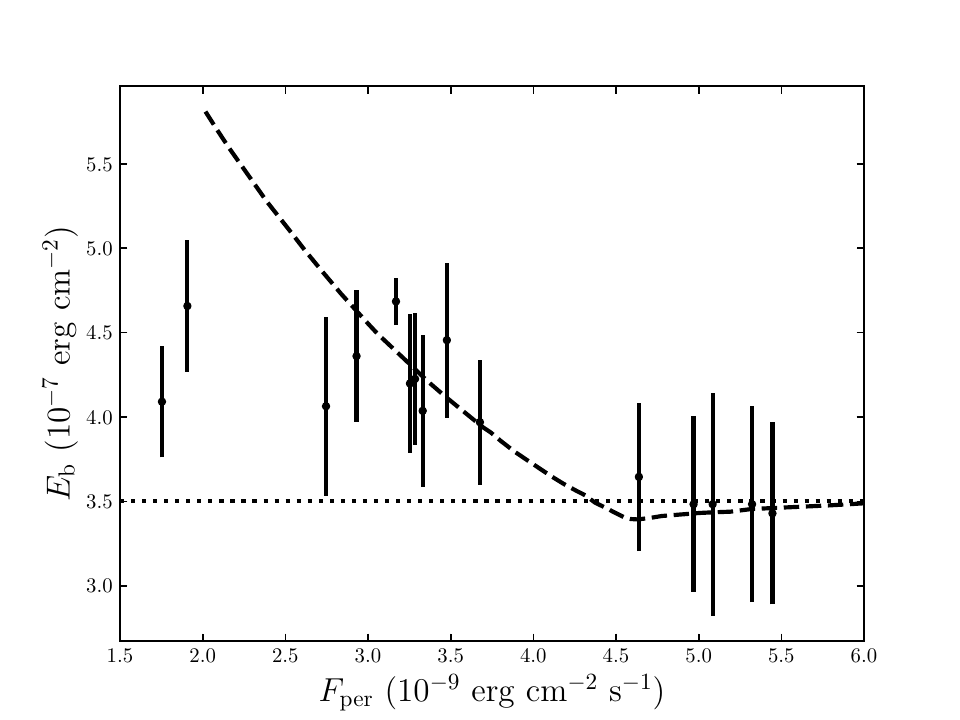}

\caption{Burst fluence vs. the persistent flux. The dotted line represents the pure helium burst, and the dashed line shows the burning of helium with hydrogen, $X=0.1$, and heavy elements, $Z=0.01$, from \citet{Cumming2003}. }
      \label{Fig:Eb_flux}
\end{figure}

The abundances of the burning matter can be inferred from the relation between the burst fluence and its persistent flux. \citet{Cumming2003} proposed that the burst fluence is independent of the persistent flux for pure helium burst. For a small fraction of hydrogen and heavy elements ($X=0.1,~Z=0.01$) in the burning material, the bursts release more energy at lower persistent flux.  
\citet{Suleimanov2017} showed that the bursts observed by \rxte\ favor the case of pure helium burning. For \nicer\ bursts, we renormalized the fluence and persistent flux to, $E_{\rm b}=3.5\times10^{-7}~{\rm erg~cm^{-2}}$, and $F_{\rm per}=4.3\times10^{-9}~{\rm erg~cm^{-2}~s^{-1}}$. In Fig.~\ref{Fig:Eb_flux}, most of the \nicer\ bursts followed the prediction of burning of helium with a mixture of hydrogen and heavy elements. However, the dependences of the burst fluence and its persistent flux in bursts \#7 and \#8 are biased from the prediction, suggesting a lower fraction of hydrogen and heavy elements in the burning material.

\begin{table*}
\begin{center} 
\caption{Overview of burst parameters.  \label{table:gap}}
\resizebox{\linewidth}{!}{\begin{tabular}{ccccccccccccc} 
\hline\\ 
{\centering Burst} &
{\centering  $\Dot{m}$  } &
{\centering  $t_{\rm se}$\tablefootmark{a}} &
{\centering  $t_{\rm me}$\tablefootmark{b}} &
{\centering  $\Delta t_\mathrm{{rec}}$\tablefootmark{c} } & 
{\centering  $\Delta T_{\mathrm{rec}}$\tablefootmark{d} } &
{\centering  $E_{\rm b}$} &
{\centering  $y_{\rm ign}$} &
{\centering  $\tau$\tablefootmark{e} } &
\\
$\#$& ($\mathrm{10^{4} ~ g ~ cm^{-2} ~ s^{-1}}$)  &(s) &(s) &(hr)&(hr)&($10^{-7}~ \mathrm{erg~cm^{-2}}$)&($10^{8}\mathrm{~g~ cm^{-2}}$)&(s)&\\ [0.01cm]\hline
1&$3.71\pm{0.53}$  &$0.8\pm{0.1}$&$2.1\pm{0.2}$&$1.4\pm{0.5}$&$-   $ &$2.58\pm{0.58}$ &$1.84\pm{0.41}$&$4.9\pm{2.4}$&\\
2&$3.79\pm{0.54}$  &         -   &-            &$1.3\pm{0.5}$&$2.2$  &$2.54\pm{0.54}$ &$1.82\pm{0.39}$&$5.0\pm{2.2}$&\\
3&$3.06\pm{0.44}$  &$0.6\pm{0.1}$&$2.5\pm{0.2}$&$1.7\pm{0.5}$&$18.3 $&$2.70\pm{0.44}$ &$1.93\pm{0.32}$&$5.7\pm{1.2}$&\\
4&$3.42\pm{0.49}$  &$0.8\pm{0.1}$&$2.3\pm{0.2}$&$1.5\pm{0.5}$&$2.8  $&$2.58\pm{0.52}$ &$1.84\pm{0.37}$&$5.2\pm{1.7}$&\\
5&$3.45\pm{0.49}$  &$0.7\pm{0.1}$&$1.9\pm{0.2}$&$1.6\pm{0.6}$&$6.3  $&$2.58\pm{0.66}$ &$1.98\pm{0.47}$&$5.2\pm{2.7}$&\\
6&$2.10\pm{0.30}$  &$1.4\pm{0.1}$&$2.6\pm{0.2}$&$3.3\pm{0.6}$&$-    $&$3.47\pm{0.14}$ &$2.47\pm{0.10}$&$6.0\pm{0.5}$&\\
7&$1.19\pm{0.17}$  &$1.2\pm{0.1}$&$3.1\pm{0.4}$&$5.0\pm{1.3}$&$133.1$&$3.03\pm{0.33}$ &$2.16\pm{0.24}$&$5.7\pm{2.0}$&\\
8&$1.21\pm{0.17}$  &$1.5\pm{0.1}$&$2.5\pm{0.2}$&$5.7\pm{1.5}$&$35.7 $&$3.45\pm{0.39}$ &$2.46\pm{0.28}$&$6.9\pm{2.0}$&\\
9&$1.93\pm{0.28}$  &$1.2\pm{0.1}$&$2.3\pm{0.2}$&$3.3\pm{0.9}$&$162.4$&$3.23\pm{0.39}$ &$2.30\pm{0.28}$&$6.9\pm{2.2}$&\\
10&$1.85\pm{0.26}$ &$1.3\pm{0.1}$&$2.3\pm{0.2}$&$2.3\pm{0.6}$&$125.3$&$3.01\pm{0.53}$ &$1.56\pm{0.15}$&$6.0\pm{1.9}$&\\
11&$2.20\pm{0.32}$ &$1.3\pm{0.1}$&$2.5\pm{0.2}$&$2.7\pm{0.8}$&$11.0 $&$2.99\pm{0.45}$ &$2.13\pm{0.32}$&$6.2\pm{2.2}$&\\
12&$2.32\pm{0.33}$ &$1.1\pm{0.1}$&$2.7\pm{0.2}$&$2.8\pm{0.8}$&$24.4 $&$3.30\pm{0.46}$ &$2.35\pm{0.33}$&$6.1\pm{2.2}$&\\
13&$2.13\pm{0.30}$ &$1.4\pm{0.1}$&$2.7\pm{0.2}$&$2.9\pm{0.8}$&$52.8 $&$3.13\pm{0.39}$ &$2.24\pm{0.28}$&$6.5\pm{1.2}$&\\
14&$2.55\pm{0.36}$ &$1.0\pm{0.1}$&$2.1\pm{0.2}$&$2.3\pm{0.6}$&$38.6 $&$2.94\pm{0.37}$ &$2.10\pm{0.27}$&$6.0\pm{0.9}$&\\
15&$2.32\pm{0.33}$ &$0.8\pm{0.1}$&$3.4\pm{0.4}$&$2.7\pm{0.7}$&$  -  $&$3.11\pm{0.41}$ &$2.22\pm{0.29}$&$6.0\pm{0.9}$&\\
\hline 
\end{tabular} }
\end{center}
\tablefoot{
\tablefoottext{a} {The duration of the superexpansion or strong photospheric expansion phase. }\\
\tablefoottext{b}{ The duration of the moderate expansion phase.}
\tablefoottext{c}{ The estimate recurrence time.}\\
\tablefoottext{d}{ The observed recurrence time. Due to data gaps, it is only considered as upper limit. }\\
\tablefoottext{e}{ The decay time of bursts defined as the ratio of the burst fluence to its peak flux.\\}
}
\end{table*}

\section{Conclusions}
\label{Sec:conclusion}
We detected 15 type I X-ray bursts in 4U 1820--30 from \nicer\ observations between 2017 and the present. The persistent spectra are well fitted by the model \texttt{TBabs} $\times$ \texttt{(bbodyrad} + \texttt{compTT)}. We adopted  the $f_{\rm a}$ model, the double-blackbody model and the reflection model \texttt{relxillNS} to fit the time-resolved burst spectra. However, for bursts \#7 and \#8, the count rates declined to the pre-burst level in 3-10 keV, which contradicts the $f_{\rm a}$ model.  The reflection model provides physically motivated and self-consistent results. From the evolution of the accretion disk density during X-ray bursts, we found direct evidence that the accretion disk was distorted by the burst radiation with a 0.5 s delay. The time-resolved burst spectra show that all bursts belong to the PRE bursts, including 13 strong photospheric expansion bursts for $r_{\rm ph}\approx100-340$ km and one superexpansion burst for $r_{\rm ph}>10^3$ km \citep{Galloway2008,int2010,keek2017}.

\begin{acknowledgements}
We appreciate the referee for the constructive and valuable comments which improves our manuscript. This work was supported by the Major Science and Technology Program of Xinjiang Uygur Autonomous Region (No. 2022A03013-3). Z.S.L. and Y.Y.P. were supported by National Natural Science Foundation of China (12103042, U1938107, 12273030, 12173103, 12122302). S.Z. is supported by National Key R\&D Program of China (2021YFA0718500). This work made use of data from the High Energy Astrophysics Science Archive Research Center (HEASARC), provided by NASA’s Goddard Space Flight Center. 

\end{acknowledgements}

\bibliography{file}{}

\begin{thebibliography}{74}
\expandafter\ifx\csname natexlab\endcsname\relax\def\natexlab#1{#1}\fi

\bibitem[{Anderson {et~al.}(1997)Anderson, Margon, Deutsch, Downes, \&
  Allen}]{Anderson_1997}
Anderson, S.~F., Margon, B., Deutsch, E.~W., Downes, R.~A., \& Allen, R.~G.
  1997, The Astrophysical Journal, 482, L69

\bibitem[{{Arnaud}(1996)}]{Arnaud96}
{Arnaud}, K.~A. 1996, in Astronomical Society of the Pacific Conference Series,
  Vol. 101, Astronomical Data Analysis Software and Systems V, ed. G.~H.
  {Jacoby} \& J.~{Barnes}, 17

\bibitem[{{Ballantyne}(2023)}]{Ballantyne2023}
{Ballantyne}, D.~R. 2023, \mnras, 518, 3357

\bibitem[{{Bilous} \& {Watts}(2019)}]{Bilous19}
{Bilous}, A.~V. \& {Watts}, A.~L. 2019, \apjs, 245, 19

\bibitem[{{Bult} {et~al.}(2021){Bult}, {Altamirano}, {Arzoumanian},
  {Ballantyne}, {Chenevez}, {Fabian}, {Gendreau}, {Homan}, {Jaisawal},
  {Malacaria}, {Miller}, {Parker}, \& {Strohmayer}}]{bult2021}
{Bult}, P., {Altamirano}, D., {Arzoumanian}, Z., {et~al.} 2021, \apj, 920, 59

\bibitem[{{Chou} \& {Grindlay}(2001)}]{Chou2001ApJ}
{Chou}, Y. \& {Grindlay}, J.~E. 2001, \apj, 563, 934

\bibitem[{{Clark} {et~al.}(1976){Clark}, {Jernigan}, {Bradt}, {Canizares},
  {Lewin}, {Li}, {Mayer}, {McClintock}, \& {Schnopper}}]{Clark76}
{Clark}, G.~W., {Jernigan}, J.~G., {Bradt}, H., {et~al.} 1976, \apjl, 207, L105

\bibitem[{{Connors} {et~al.}(2020){Connors}, {Garc{\'\i}a}, {Dauser},
  {Grinberg}, {Steiner}, {Sridhar}, {Wilms}, {Tomsick}, {Harrison}, \&
  {Licklederer}}]{Connors20}
{Connors}, R. M.~T., {Garc{\'\i}a}, J.~A., {Dauser}, T., {et~al.} 2020, \apj,
  892, 47

\bibitem[{{Costantini} {et~al.}(2012){Costantini}, {Pinto}, {Kaastra}, {in't
  Zand}, {Freyberg}, {Kuiper}, {M{\'e}ndez}, {de Vries}, \&
  {Waters}}]{Costantini2012}
{Costantini}, E., {Pinto}, C., {Kaastra}, J.~S., {et~al.} 2012, \aap, 539, A32

\bibitem[{{Cumming}(2003)}]{Cumming2003}
{Cumming}, A. 2003, \apj, 595, 1077

\bibitem[{{Cumming} \& {Macbeth}(2004)}]{Cumming2004ApJ}
{Cumming}, A. \& {Macbeth}, J. 2004, \apjl, 603, L37

\bibitem[{{Cumming} {et~al.}(2006){Cumming}, {Macbeth}, {in 't Zand}, \&
  {Page}}]{Cumming2006ApJ}
{Cumming}, A., {Macbeth}, J., {in 't Zand}, J.~J.~M., \& {Page}, D. 2006, \apj,
  646, 429

\bibitem[{Degenaar {et~al.}(2018)Degenaar, Ballantyne, Belloni, Chakraborty,
  Chen, Ji, Kretschmar, Kuulkers, Li, Maccarone, Malzac, Zhang, \&
  Zhang}]{Nathalie2018Accretion}
Degenaar, N., Ballantyne, D.~R., Belloni, T., {et~al.} 2018, SPACE SCIENCE
  REVIEWS, 214

\bibitem[{{Degenaar} {et~al.}(2018){Degenaar}, {Ballantyne}, {Belloni},
  {Chakraborty}, {Chen}, {Ji}, {Kretschmar}, {Kuulkers}, {Li}, {Maccarone},
  {Malzac}, {Zhang}, \& {Zhang}}]{Degenaar18}
{Degenaar}, N., {Ballantyne}, D.~R., {Belloni}, T., {et~al.} 2018, \ssr, 214,
  15

\bibitem[{{Degenaar} {et~al.}(2016){Degenaar}, {Koljonen}, {Chakrabarty},
  {Kara}, {Altamirano}, {Miller}, \& {Fabian}}]{Degenaar16}
{Degenaar}, N., {Koljonen}, K.~I.~I., {Chakrabarty}, D., {et~al.} 2016, \mnras,
  456, 4256

\bibitem[{{Falanga} {et~al.}(2008){Falanga}, {Chenevez}, {Cumming}, {Kuulkers},
  {Trap}, \& {Goldwurm}}]{Falanga2008A&A}
{Falanga}, M., {Chenevez}, J., {Cumming}, A., {et~al.} 2008, \aap, 484, 43

\bibitem[{{Fragile} {et~al.}(2020){Fragile}, {Ballantyne}, \&
  {Blankenship}}]{Fragile2020}
{Fragile}, P.~C., {Ballantyne}, D.~R., \& {Blankenship}, A. 2020, Nature
  Astronomy, 4, 541

\bibitem[{{Galloway} \& {Keek}(2021)}]{Galloway21}
{Galloway}, D.~K. \& {Keek}, L. 2021, Astrophysics and Space Science Library,
  461, 209

\bibitem[{{Galloway} {et~al.}(2008{\natexlab{a}}){Galloway}, {Muno}, {Hartman},
  {Psaltis}, \& {Chakrabarty}}]{Galloway08}
{Galloway}, D.~K., {Muno}, M.~P., {Hartman}, J.~M., {Psaltis}, D., \&
  {Chakrabarty}, D. 2008{\natexlab{a}}, \apjs, 179, 360

\bibitem[{{Galloway} {et~al.}(2008{\natexlab{b}}){Galloway}, {Muno}, {Hartman},
  {Psaltis}, \& {Chakrabarty}}]{Galloway2008}
{Galloway}, D.~K., {Muno}, M.~P., {Hartman}, J.~M., {Psaltis}, D., \&
  {Chakrabarty}, D. 2008{\natexlab{b}}, \apjs, 179, 360

\bibitem[{{Garc{\'\i}a} {et~al.}(2013){Garc{\'\i}a}, {Zhang}, \&
  {M{\'e}ndez}}]{garc2013}
{Garc{\'\i}a}, F., {Zhang}, G., \& {M{\'e}ndez}, M. 2013, \mnras, 429, 3266

\bibitem[{{Giacconi} {et~al.}(1974){Giacconi}, {Murray}, {Gursky}, {Kellogg},
  {Schreier}, {Matilsky}, {Koch}, \& {Tananbaum}}]{uhuru_1974ApJS}
{Giacconi}, R., {Murray}, S., {Gursky}, H., {et~al.} 1974, \apjs, 27, 37

\bibitem[{{Goodwin} {et~al.}(2019){Goodwin}, {Heger}, \&
  {Galloway}}]{Goodwin19}
{Goodwin}, A.~J., {Heger}, A., \& {Galloway}, D.~K. 2019, \apj, 870, 64

\bibitem[{{Guichandut} {et~al.}(2021){Guichandut}, {Cumming}, {Falanga}, {Li},
  \& {Zamfir}}]{Guichandut2021ApJ}
{Guichandut}, S., {Cumming}, A., {Falanga}, M., {Li}, Z., \& {Zamfir}, M. 2021,
  \apj, 914, 49

\bibitem[{{G{\"u}ver} {et~al.}(2010){G{\"u}ver}, {Wroblewski}, {Camarota}, \&
  {{\"O}zel}}]{Guver2010}
{G{\"u}ver}, T., {Wroblewski}, P., {Camarota}, L., \& {{\"O}zel}, F. 2010,
  \apj, 719, 1807

\bibitem[{Güver {et~al.}(2021)Güver, Boztepe, Ballantyne, Bostancı, Bult,
  Jaisawal, Göğüş, Strohmayer, Altamirano, Guillot, \&
  Chakrabarty}]{Guver22}
Güver, T., Boztepe, T., Ballantyne, D.~R., {et~al.} 2021, Monthly Notices of
  the Royal Astronomical Society, 510, 1577

\bibitem[{{He} \& {Keek}(2016)}]{He16}
{He}, C.~C. \& {Keek}, L. 2016, \apj, 819, 47

\bibitem[{{in 't Zand}(2011)}]{zand2011arXiv}
{in 't Zand}, J. 2011, arXiv e-prints, arXiv:1102.3345

\bibitem[{in~'t Zand(2017)}]{int2017}
in~'t Zand, J. 2017, Understanding superbursts

\bibitem[{{in't Zand} {et~al.}(2011{\natexlab{a}}){in't Zand}, {Serino},
  {Kawai}, \& {Heinke}}]{int2011ATel}
{in't Zand}, J., {Serino}, M., {Kawai}, N., \& {Heinke}, C. 2011{\natexlab{a}},
  The Astronomer's Telegram, 3625, 1

\bibitem[{{in't Zand} {et~al.}(2014){in't Zand}, {Cumming}, {Triemstra},
  {Mateijsen}, \& {Bagnoli}}]{int2014A&A}
{in't Zand}, J.~J.~M., {Cumming}, A., {Triemstra}, T.~L., {Mateijsen},
  R.~A.~D.~A., \& {Bagnoli}, T. 2014, \aap, 562, A16

\bibitem[{{in't Zand} {et~al.}(2011{\natexlab{b}}){in't Zand}, {Galloway}, \&
  {Ballantyne}}]{int2011A&A}
{in't Zand}, J.~J.~M., {Galloway}, D.~K., \& {Ballantyne}, D.~R.
  2011{\natexlab{b}}, \aap, 525, A111

\bibitem[{{in't Zand} {et~al.}(2013){in't Zand}, {Galloway}, {Marshall},
  {Ballantyne}, {Jonker}, {Paerels}, {Palmer}, {Patruno}, \&
  {Weinberg}}]{Zand13}
{in't Zand}, J.~J.~M., {Galloway}, D.~K., {Marshall}, H.~L., {et~al.} 2013,
  \aap, 553, A83

\bibitem[{{in't Zand} {et~al.}(2012){in't Zand}, {Homan}, {Keek}, \&
  {Palmer}}]{int2012A&A}
{in't Zand}, J.~J.~M., {Homan}, J., {Keek}, L., \& {Palmer}, D.~M. 2012, \aap,
  547, A47

\bibitem[{{in't Zand} {et~al.}(2019){in't Zand}, {Kries}, {Palmer}, \&
  {Degenaar}}]{int2019A&A}
{in't Zand}, J.~J.~M., {Kries}, M.~J.~W., {Palmer}, D.~M., \& {Degenaar}, N.
  2019, \aap, 621, A53

\bibitem[{{in't Zand} \& {Weinberg}(2010)}]{int2010}
{in't Zand}, J.~J.~M. \& {Weinberg}, N.~N. 2010, \aap, 520, A81

\bibitem[{{Jager} {et~al.}(1997){Jager}, {Mels}, {Brinkman}, {Galama},
  {Goulooze}, {Heise}, {Lowes}, {Muller}, {Naber}, {Rook}, {Schuurhof},
  {Schuurmans}, \& {Wiersma}}]{Jager97}
{Jager}, R., {Mels}, W.~A., {Brinkman}, A.~C., {et~al.} 1997, \aaps, 125, 557

\bibitem[{{Jahoda} {et~al.}(2006){Jahoda}, {Markwardt}, {Radeva}, {Rots},
  {Stark}, {Swank}, {Strohmayer}, \& {Zhang}}]{Jahoda06}
{Jahoda}, K., {Markwardt}, C.~B., {Radeva}, Y., {et~al.} 2006, \apjs, 163, 401

\bibitem[{{Kajava} {et~al.}(2017){Kajava}, {Koljonen}, {N{\"a}ttil{\"a}},
  {Suleimanov}, \& {Poutanen}}]{Kajava17}
{Kajava}, J.~J.~E., {Koljonen}, K.~I.~I., {N{\"a}ttil{\"a}}, J., {Suleimanov},
  V., \& {Poutanen}, J. 2017, \mnras, 472, 78

\bibitem[{{Keek}(2012)}]{keek2012ApJ}
{Keek}, L. 2012, \apj, 756, 130

\bibitem[{{Keek} {et~al.}(2018){Keek}, {Arzoumanian}, {Chakrabarty},
  {Chenevez}, {Gendreau}, {Guillot}, {G{\"u}ver}, {Homan}, {Jaisawal},
  {LaMarr}, {Lamb}, {Mahmoodifar}, {Markwardt}, {Okajima}, {Strohmayer}, \& {in
  't Zand}}]{Keek2018}
{Keek}, L., {Arzoumanian}, Z., {Chakrabarty}, D., {et~al.} 2018, \apjl, 856,
  L37

\bibitem[{{Keek} {et~al.}(2015){Keek}, {Cumming}, {Wolf}, {Ballantyne},
  {Suleimanov}, {Kuulkers}, \& {Strohmayer}}]{keek2015}
{Keek}, L., {Cumming}, A., {Wolf}, Z., {et~al.} 2015, \mnras, 454, 3559

\bibitem[{{Keek} {et~al.}(2017){Keek}, {Iwakiri}, {Serino}, {Ballantyne}, {in't
  Zand}, \& {Strohmayer}}]{keek2017}
{Keek}, L., {Iwakiri}, W., {Serino}, M., {et~al.} 2017, \apj, 836, 111

\bibitem[{{Koljonen} {et~al.}(2016){Koljonen}, {Kajava}, \&
  {Kuulkers}}]{Koljonen16}
{Koljonen}, K.~I.~I., {Kajava}, J.~J.~E., \& {Kuulkers}, E. 2016, \apj, 829, 91

\bibitem[{{Leahy} {et~al.}(1983){Leahy}, {Darbro}, {Elsner}, {Weisskopf},
  {Sutherland}, {Kahn}, \& {Grindlay}}]{Leahy83}
{Leahy}, D.~A., {Darbro}, W., {Elsner}, R.~F., {et~al.} 1983, \apj, 266, 160

\bibitem[{{Lewin} {et~al.}(1993){Lewin}, {van Paradijs}, \& {Taam}}]{Lewin93}
{Lewin}, W.~H.~G., {van Paradijs}, J., \& {Taam}, R.~E. 1993, \ssr, 62, 223

\bibitem[{{Li} {et~al.}(2021){Li}, {Pan}, \& {Falanga}}]{li2021ApJ}
{Li}, Z., {Pan}, Y., \& {Falanga}, M. 2021, \apj, 920, 35

\bibitem[{{Li} {et~al.}(2022){Li}, {Yu}, {Lu}, {Pan}, \& {Falanga}}]{li2022ApJ}
{Li}, Z., {Yu}, W., {Lu}, Y., {Pan}, Y., \& {Falanga}, M. 2022, \apj, 935, 123

\bibitem[{{Lu} {et~al.}(2023){Lu}, {Li}, {Pan}, {Yu}, {Chen}, {Ji}, {Ge},
  {Zhang}, {Qu}, {Song}, \& {Falanga}}]{Lu2023A&A}
{Lu}, Y., {Li}, Z., {Pan}, Y., {et~al.} 2023, \aap, 670, A87

\bibitem[{{Ludlam} {et~al.}(2019){Ludlam}, {Miller}, {Barret}, {Cackett},
  {Coughenour}, {Dauser}, {Degenaar}, {Garc{\'\i}a}, {Harrison}, \&
  {Paerels}}]{Ludlam19}
{Ludlam}, R.~M., {Miller}, J.~M., {Barret}, D., {et~al.} 2019, \apj, 873, 99

\bibitem[{{Nelson} {et~al.}(1986){Nelson}, {Rappaport}, \&
  {Joss}}]{Nelson1986ApJ}
{Nelson}, L.~A., {Rappaport}, S.~A., \& {Joss}, P.~C. 1986, \apj, 304, 231

\bibitem[{{Paczynski} \& {Proszynski}(1986)}]{Paczynski1986ApJ}
{Paczynski}, B. \& {Proszynski}, M. 1986, \apj, 302, 519

\bibitem[{{Paczynski} \& {Sienkiewicz}(1981)}]{1981Paczynski}
{Paczynski}, B. \& {Sienkiewicz}, R. 1981, \apjl, 248, L27

\bibitem[{{Rappaport} {et~al.}(1982){Rappaport}, {Joss}, \&
  {Webbink}}]{Rappaport1982ApJ}
{Rappaport}, S., {Joss}, P.~C., \& {Webbink}, R.~F. 1982, \apj, 254, 616

\bibitem[{{Rappaport} {et~al.}(1987){Rappaport}, {Nelson}, {Ma}, \&
  {Joss}}]{Rappaport1987}
{Rappaport}, S., {Nelson}, L.~A., {Ma}, C.~P., \& {Joss}, P.~C. 1987, \apj,
  322, 842

\bibitem[{{Remillard} {et~al.}(2021){Remillard}, {Loewenstein}, {Steiner},
  {Prigozhin}, {LaMarr}, {Enoto}, {Gendreau}, {Arzoumanian}, {Markwardt},
  {Basak}, {Stevens}, {Ray}, {Altamirano}, \& {Buisson}}]{Remillard21}
{Remillard}, R.~A., {Loewenstein}, M., {Steiner}, J.~F., {et~al.} 2021, arXiv
  e-prints, arXiv:2105.09901

\bibitem[{{Serino} {et~al.}(2021{\natexlab{a}}){Serino}, {Iwakiri}, {Negoro},
  {Nakajima}, {Kobayashi}, {Asakura}, {Seino}, {Mihara}, {Tamagawa},
  {Matsuoka}, {Sakamoto}, {Sugita}, {Komachi}, {Hiramatsu}, {Yoshida},
  {Tsuboi}, {Kawai}, {Okamoto}, {Kitakoga}, {Shidatsu}, {Kawai}, {Niwano},
  {Hosokawa}, {Imai}, {Ito}, {Takamatsu}, {Nakahira}, {Ueno}, {Tomida},
  {Ishikawa}, {Tominaga}, {Nagatsuka}, {Kurihara}, {Ueda}, {Yamada}, {Ogawa},
  {Setoguchi}, {Yoshitake}, {Goto}, {Uematsu}, {Inaba}, {Tsunemi}, {Yamauchi},
  {Nonaka}, {Sato}, {Hatsuda}, {Fukuoka}, {Kawamuro}, {Yamaoka}, {Kawakubo}, \&
  {Sugizaki}}]{b2021ATel}
{Serino}, M., {Iwakiri}, W., {Negoro}, H., {et~al.} 2021{\natexlab{a}}, The
  Astronomer's Telegram, 15071, 1

\bibitem[{{Serino} {et~al.}(2021{\natexlab{b}}){Serino}, {Iwakiri}, {Negoro},
  {Nakajima}, {Kobayashi}, {Asakura}, {Seino}, {Mihara}, {Tamagawa},
  {Matsuoka}, {Sakamoto}, {Sugita}, {Komachi}, {Yoshida}, {Tsuboi}, {Kawai},
  {Okamoto}, {Kitakoga}, {Shidatsu}, {Kawai}, {Niwano}, {Hosokawa}, {Nakahira},
  {Sugawara}, {Ueno}, {Tomida}, {Ishikawa}, {Tominaga}, {Nagatsuka}, {Ueda},
  {Yamada}, {Ogawa}, {Setoguchi}, {Yoshitake}, {Goto}, {Uematsu}, {Tsunemi},
  {Yamauchi}, {Nonaka}, {Sato}, {Hatsuda}, {Fukuoka}, {Kawamuro}, {Yamaoka},
  {Kawakubo}, \& {Sugizaki}}]{a2021ATel}
{Serino}, M., {Iwakiri}, W., {Negoro}, H., {et~al.} 2021{\natexlab{b}}, The
  Astronomer's Telegram, 14871, 1

\bibitem[{{Serino} {et~al.}(2016){Serino}, {Iwakiri}, {Tamagawa}, {Sakamoto},
  {Nakahira}, {Matsuoka}, {Yamaoka}, \& {Negoro}}]{Serino2016}
{Serino}, M., {Iwakiri}, W., {Tamagawa}, T., {et~al.} 2016, \pasj, 68, 95

\bibitem[{{Speicher} {et~al.}(2022){Speicher}, {Ballantyne}, \&
  {Fragile}}]{Speicher2022}
{Speicher}, J., {Ballantyne}, D.~R., \& {Fragile}, P.~C. 2022, \mnras, 509,
  1736

\bibitem[{{Stella} {et~al.}(1987){Stella}, {Priedhorsky}, \&
  {White}}]{Stella1987}
{Stella}, L., {Priedhorsky}, W., \& {White}, N.~E. 1987, \apjl, 312, L17

\bibitem[{{Strohmayer} \& {Bildsten}(2006)}]{Strohmayer06}
{Strohmayer}, T. \& {Bildsten}, L. 2006, in Compact stellar X-ray sources,
  Cambridge Astrophysics Series, No. 39, ed. W.~{Lewin} \& M.~{van der Klis}
  (Cambridge: Cambridge University Press), 113--156

\bibitem[{{Strohmayer} {et~al.}(2019){Strohmayer}, {Altamirano}, {Arzoumanian},
  {Bult}, {Chakrabarty}, {Chenevez}, {Fabian}, {Gendreau}, {Guillot}, {in 't
  Zand}, {Jaisawal}, {Keek}, {Kosec}, {Ludlam}, {Mahmoodifar}, {Malacaria}, \&
  {Miller}}]{Strohmayer2019}
{Strohmayer}, T.~E., {Altamirano}, D., {Arzoumanian}, Z., {et~al.} 2019, \apjl,
  878, L27

\bibitem[{Strohmayer \& Brown(2002)}]{Strohmayer_2002}
Strohmayer, T.~E. \& Brown, E.~F. 2002, The Astrophysical Journal, 566, 1045

\bibitem[{{Suleimanov} {et~al.}(2017){Suleimanov}, {Kajava}, {Molkov},
  {N{\"a}ttil{\"a}}, {Lutovinov}, {Werner}, \& {Poutanen}}]{Suleimanov2017}
{Suleimanov}, V.~F., {Kajava}, J. J.~E., {Molkov}, S.~V., {et~al.} 2017,
  \mnras, 472, 3905

\bibitem[{{Tawara} {et~al.}(1984){Tawara}, {Hayakawa}, \& {Kii}}]{Tawara1984}
{Tawara}, Y., {Hayakawa}, S., \& {Kii}, T. 1984, \pasj, 36, 845

\bibitem[{{Titarchuk}(1994)}]{Lev}
{Titarchuk}, L. 1994, \apj, 434, 570

\bibitem[{{Valenti} {et~al.}(2004){Valenti}, {Ferraro}, \&
  {Origlia}}]{Valenti2004}
{Valenti}, E., {Ferraro}, F.~R., \& {Origlia}, L. 2004, \mnras, 351, 1204

\bibitem[{{Verbunt} \& {van den Heuvel}(1995)}]{1995xrbi.nasa..457V}
{Verbunt}, F. \& {van den Heuvel}, E.~P.~J. 1995, in X-ray Binaries, 457--494

\bibitem[{White \& Zhang(1997)}]{White_1997}
White, N.~E. \& Zhang, W. 1997, The Astrophysical Journal, 490, L87

\bibitem[{Wilms {et~al.}(2000)Wilms, Allen, \& McCray}]{Wilms_2000}
Wilms, J., Allen, A., \& McCray, R. 2000, The Astrophysical Journal, 542, 914

\bibitem[{{Worpel} {et~al.}(2013){Worpel}, {Galloway}, \& {Price}}]{Worpel13}
{Worpel}, H., {Galloway}, D.~K., \& {Price}, D.~J. 2013, \apj, 772, 94

\bibitem[{{Yu} \& {Weinberg}(2018)}]{Yu2018}
{Yu}, H. \& {Weinberg}, N.~N. 2018, \apj, 863, 53

\bibitem[{{Zhao} {et~al.}(2022){Zhao}, {Li}, {Pan}, {Falanga}, {Ji}, {Chen}, \&
  {Zhang}}]{Zhao22}
{Zhao}, G., {Li}, Z., {Pan}, Y., {et~al.} 2022, \aap, 660, A31

\end{thebibliography}
\bibliographystyle{aa}

\begin{appendix} %
\onecolumn
\section{Figure} \label{Sec:App_Table}
\centering

\begin{figure*}[h!]
    
    \centering
        \includegraphics[width=\hsize]{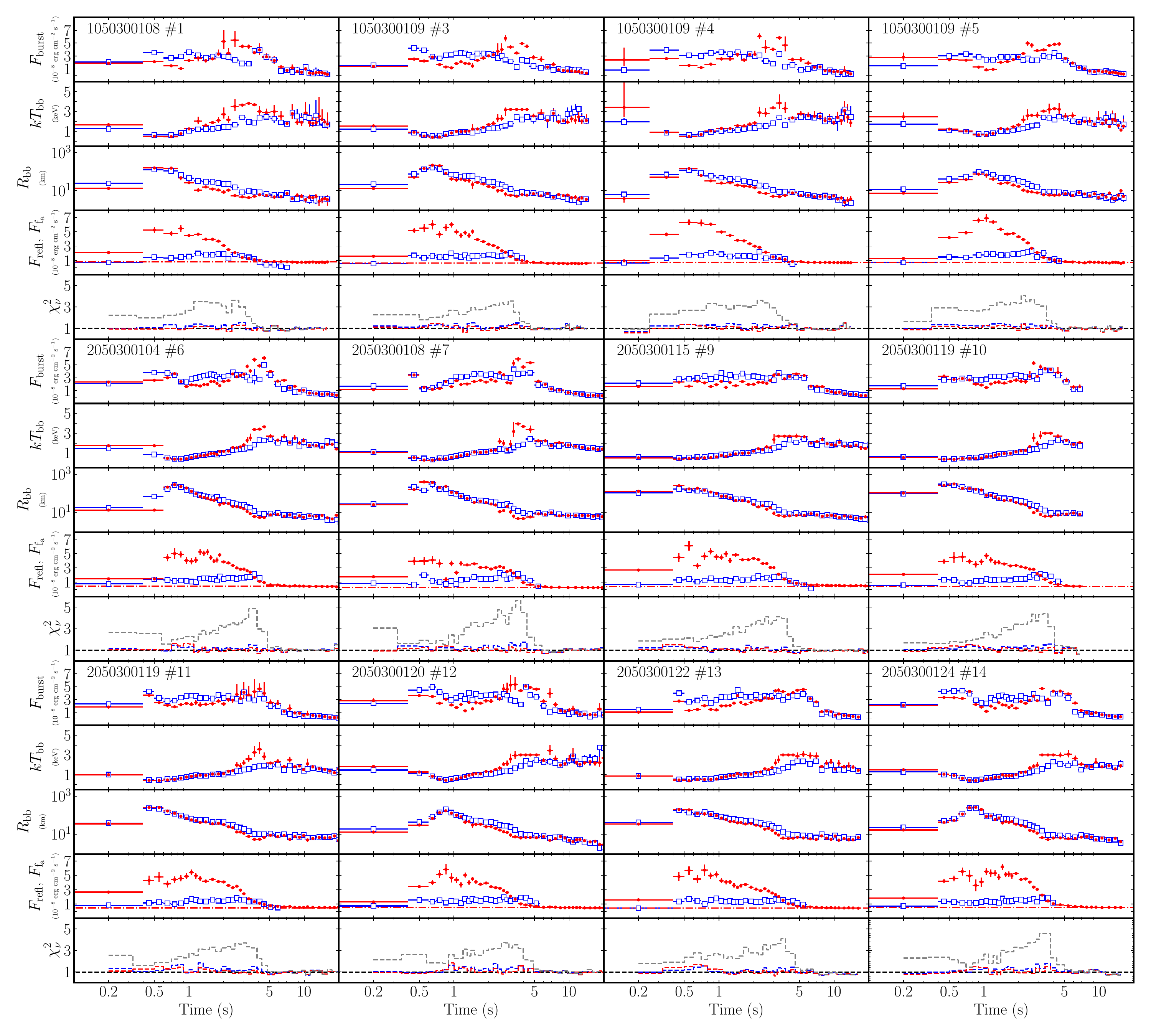}
                \caption{Same as Fig.~\ref{Fig:fa_re}, but for the rest of  bursts.} 
        \label{Fig:fa_re_all}%
\end{figure*}  

\begin{figure*}
    
    \centering
        \includegraphics[width=\hsize]{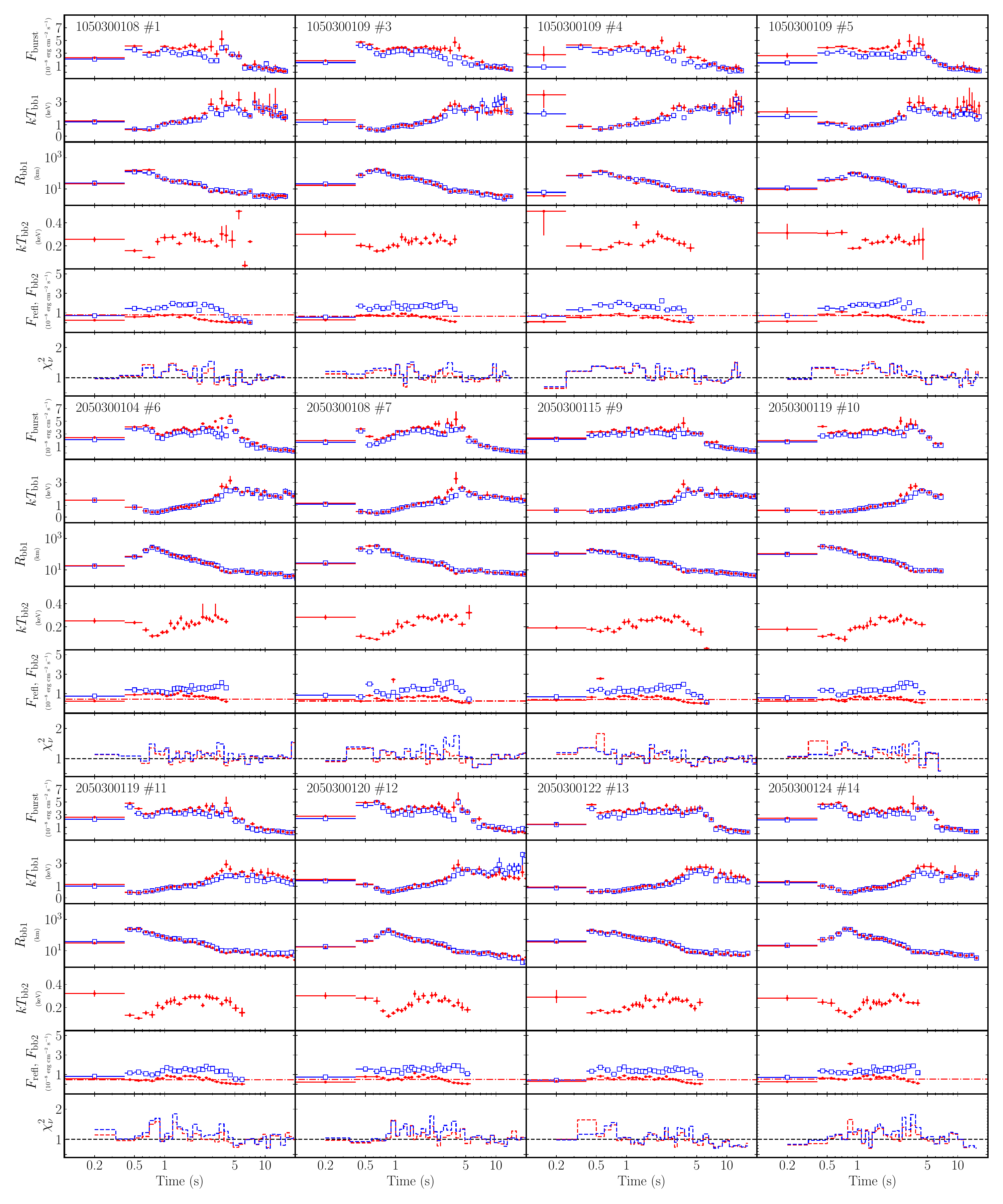}
        \caption{Same as Fig.~\ref{Fig:2bb}, but for the rest of bursts.} 
        \label{Fig:2bb_all}%
\end{figure*}

\end{appendix}

\end{document}